\documentclass[useAMS,usenatbib,referee]{biom}

\usepackage{bm}
\usepackage{amssymb,mathtools}
\usepackage{natbib}
\usepackage[hyphens]{url}
\usepackage[linesnumbered, ruled]{algorithm2e}
\usepackage{hyperref}
\hypersetup{
colorlinks=true,
linkcolor=red,
filecolor=magenta,
citecolor=blue,
urlcolor=blue,
}
\usepackage{cleveref}
\Crefname{assumption}{Assumption}{Assumptions}
\usepackage{setspace}
\usepackage{tikz}
\usepackage{calc} 
\tikzset{>=latex} 

\usepackage{bbm}
\usepackage{textcomp} 

\usepackage{multirow,multicol,makecell,booktabs}

\def\bbeta{\bm{\beta}}

\def\argmax{\operatornamewithlimits{argmax}}

\usepackage{subfigure}

\title[]{Bregman Divergence-Based Data Integration with Application to Polygenic Risk Score (PRS) Heterogeneity Adjustment}

\author{Qinmengge Li $^{1}$, Matthew T. Patrick $^{2}$, Haihan Zhang $^{1}$, Chachrit Khunsriraksakul $^{3}$, \newauthorline Philip E. Stuart $^{2}$, Johann E. Gudjonsson $^{2}$, Rajan Nair$^{2}$, James T. Elder $^{2}$, Dajiang J. Liu $^{3}$, \newauthorline Jian Kang $^{1}$, Lam C. Tsoi $^{2}$, Kevin He $^{1,*}$\email{kevinhe@umich.edu}\\$^1$Department of Biostatistics, University of Michigan, Ann Arbor, MI 48109,\\$^2$Department of Dermatology, University of Michigan, Ann Arbor, MI 48109,\\$^3$College of Medicine, Pennsylvania State University, PA 16802}

\begin{document}

\pagerange{\pageref{firstpage}--\pageref{lastpage}} 

\label{firstpage}

\begin{abstract}
Polygenic risk scores (PRS), developed as the sum of single-nucleotide polymorphisms (SNPs) weighted by the risk allele effect sizes as estimated by published genome-wide association studies, have recently received much attention for genetics risk prediction. While successful for the Caucasian population, the PRS based on the minority population cohorts suffer from limited event rates, small sample sizes, high dimensionality and low signal-to-noise ratios, exacerbating already severe health disparities.  Due to population heterogeneity, direct trans-ethnic prediction by utilizing the Caucasian model for the minority population also has limited performance. As a result, it is desirable to design a data integration procedure to measure the difference between the populations and optimally balance the information from them to improve the prediction stability of the minority populations. A unique challenge here is that due to data privacy, the individual genotype data is not accessible for either the Caucasian population or the minority population. Therefore, new data integration methods based only on encrypted summary statistics are needed. To address these challenges, we propose a BRegman divergence-based Integrational Genetic Hazard Trans-ethnic (BRIGHT) estimation procedure to transfer the information we learned from PRS across different ancestries. The proposed method only requires the use of published summary statistics and can be applied to improve the performance of PRS for ethnic minority groups, accounting for challenges including potential model misspecification, data heterogeneity and data sharing constraints. We provide the asymptotic consistency and weak oracle property for the proposed method. Simulations show the prediction and variable selection advantages of the proposed method when applied to heterogeneous datasets. Real data analysis constructing psoriasis PRS scores for a South Asian population also confirms the improved model performance.
\end{abstract}


\maketitle

\section{Introduction}

Genetic risk prediction is an important topic in the study of precision health and preventive medicine \citep{torkamani2018personal, chatterjee2016developing}. However, the very large number of features (e.g. millions of SNPs) relative to the small sample sizes makes it extremely difficult to construct robust prediction models. Moreover, because patient identities can be fully recovered by utilizing the genotype information  \citep{wang2016individual}, privacy-preserving is crucial, making individual-level genotypes hard to access and share \citep{ozair2015ethical}. 
 
In recent years, great efforts \citep{chatterjee2013projecting,dudbridge2013power,shah2015improving, vilhjalmsson2015modeling} have been made with polygenic risk scores (PRS), which summarize the genetic contributions to a disease or phenotype based on published genome-wide association studies (GWAS). For instance, the LASSOsum method \citep{mak2017polygenic} utilized the LASSO regression  \citep{tibshirani1996regression} to select important markers using summary-level information only. Because only summary statistics are required, the algorithm preserves the privacy of patients, and therefore achieves greater flexibility in data sharing and integrative analysis. 

Despite the potential usefulness of PRS, one major limitation hindering its broad application is that little work has been done for minority populations.
Based on a recent work by \cite{duncan2019analysis}, 67\% and 19\% of the genetic risk studies during 2008-2017 were exclusively performed on Caucasian and East Asian populations, respectively, and only 3.8\% investigated other populations, raising concerns for equitable precision medicine accessibility \citep{ martin2017human,wang2020theoretical}. Indeed, constructing PRS for minority populations suffers from limited sample sizes, with no guarantee of prediction accuracy \citep{marquez2017multiethnic, ruan2021improving}. A desirable solution is to transfer knowledge across ethnicities to improve the prediction performance for minority populations. 
 
Existing trans-ethnic predictions were typically developed under a strong assumption that the genetic structures underlying complex traits are homogeneous across ethnicities. However, this assumption is often violated, in that while the same genes and their expressed proteins are likely to participate in the disease process across ethnicities, the genetic evidence pointing to their involvement may differ greatly across diverse human populations. Recent studies have shown that the performance of directly applying the PRS constructed on the majority population to target minority data is limited, especially when studying African American and South Asian populations \citep{duncan2019analysis, vilhjalmsson2015modeling}, which is known as the lack of model transferability. The reduced prediction power is due to heterogeneous genotype (predictors) structures including different minor allele frequencies (MAF) and linkage disequilibrium (LD) patterns, as well as unobserved environmental effects leading to distinct genotype-trait associations across different ethnic populations \citep{vilhjalmsson2015modeling}. Alternatively, \cite{marquez2017multiethnic} proposed weighting traditional PRS across different ethnicities to maximize their performance in the target population. However, individual-level data are required for tuning the optimal weights and, hence, the resulting algorithm violates the privacy-preserving requirements.
 
Our proposed solution was motivated by divergence measures, originally developed in the fields of information theory to measure the difference between the two probability distributions. In particular, \cite{liu2003kullback} and \cite{schapire2005boosting} proposed using Kullback–Leibler divergence \citep{kullback1951information} to integrate multiple data sources within the  boosting framework. \cite{jiang2016variable} applied the KL divergence to the LASSO regression for variable selection. \cite{wang2021kullback} extended the KL divergence to discrete failure time models. However, it is noteworthy that the KL-based integration procedures rely on a measure of the discrepancy between two probability distributions, such as the Gaussian distribution, for which the model assumption is often violated in genetics studies. To relax the restricted assumption, a more robust integration procedure is needed. More importantly, due to data privacy, a unique challenge for the trans-ethnic PRS is that the individual-level data are not accessible in either the Caucasian population or in other minority populations, precluding the applications of the aforementioned KL-based integration procedures. 

To fill these gaps, we propose a novel Bregman divergence-based data integration procedure. Our overall goal is to borrow the knowledge learned from well-established Caucasian genetic risk prediction models to improve the prediction performance in a minority population, accounting for challenges such as potential model misspecification, data heterogeneity and data sharing constraints. Specifically, to measure the fit of the prediction models to the minority population, we utilize the modified Ordinary Least Square (OLS) loss.  
To measure the heterogeneity across ethnicity, we propose using a Bregman divergence to measure the heterogeneity between the data sources and optimally balance the information between them.
The contributions of the proposed method can be summarized in the following aspects: (1) it controls the relative weight of external risk prediction models, identifying the most compatible ones and diminishing the weights of less relevant ones; (2) only summary statistics are required with no use of any individual-level data, hence preserving patient privacy; (3) the procedure is robust to model misspecifications and incorporates the KL-divergence as a special case; (4) it accommodates that external prediction models may only include partial information on a subset of the predictors in the target minority population; and (5) the resulting integration procedure retain similar forms as the classical penalized regressions such as the LASSO, and hence, the corresponding estimation is computationally efficient and can be easily applied to ultrahigh-dimensional genetics risk predictions.

The rest of the paper is organized as follows: In Section \ref{Method}, we introduce the prerequisite notations and then propose the Bregman divergence-based data integration framework. Section \ref{Theoretical properties} provides the asymptotic consistency and weak oracle property \cite{fan2010selective} for the proposed method. Section \ref{simulation} conducts simulations to evaluate the proposed method. In Section \ref{Real data}, a data analysis for psoriasis risk prediction is presented. Discussions are summarized in Section \ref{discussion}.

\section{Methods} \label{Method}

\subsection{Notation and model for the target minority population} \label{Notations and model assumptions}

For the target minority population, consider a linear regression with $n$  independent samples, 
\begin{align}
    \boldsymbol y=\boldsymbol X\boldsymbol\beta_0+\boldsymbol\epsilon,\label{eq: underlying model}
\end{align}
where $\boldsymbol X$ is an $n\times p$ genotype covariate matrix,  $\boldsymbol y$ is a $n$-dimensional response vector, $\boldsymbol\beta_0$ is the corresponding $p$-dimensional coefficient vector, and $\boldsymbol\epsilon$ is a $n$-dimensional vector of independently and identically distributed random errors with $E(\boldsymbol\epsilon)=\boldsymbol 0$.

Because $p \gg n$,  $\boldsymbol\beta$ is difficult to estimate without the common sparsity condition, which is particularly relevant in genetics studies, for which typically only a relatively small number of SNPs are related to the response. For  improved risk prediction and model interpretability, we aim to identify the active set
$\mathcal{S}_0=\{j: \beta_{0,j}\neq0, ~j=1,\ldots, p\}.$

If both $\boldsymbol X$ and $\boldsymbol y$ were available, classical penalized regressions, such as LASSO, could be applied. However, due to patient privacy requirements, the individual-level design matrix $\boldsymbol X$ and outcome vector $\boldsymbol y$ are usually not available for PRS construction. 
Instead, only the de-identified GWAS summary information (e.g. $\boldsymbol X^\top \boldsymbol y$) can be accessed. Further details for GWAS summary statistics are provided in Supplementary Materials. 

\subsection{Notation for prior information}

Due to privacy-preserving requirements, individual-level data for the Caucasian population are generally not accessible. Instead, only coefficient estimates $\boldsymbol{\tilde\beta}$ are publicly available, which is encrypted and information leakage free. 
Two examples of $\boldsymbol{\tilde\beta}$ are given as follows: 

{\it Example 1}: When single-ethnic PRS estimates are available for the Caucasian population, these estimates can directly be used as $\boldsymbol{\tilde\beta}$.

{\it Example 2}: When GWAS summary statistics are available for the Caucasian population,  $\boldsymbol{\tilde\beta}$ can be generated from a well-established single-ethnic PRS method, such as the LASSOsum.

\subsection{BRIGHT estimation procedure} \label{BRIGHT estimation procedure}
\begin{figure}[h]
    \centering
    \includegraphics[width=0.99\textwidth]{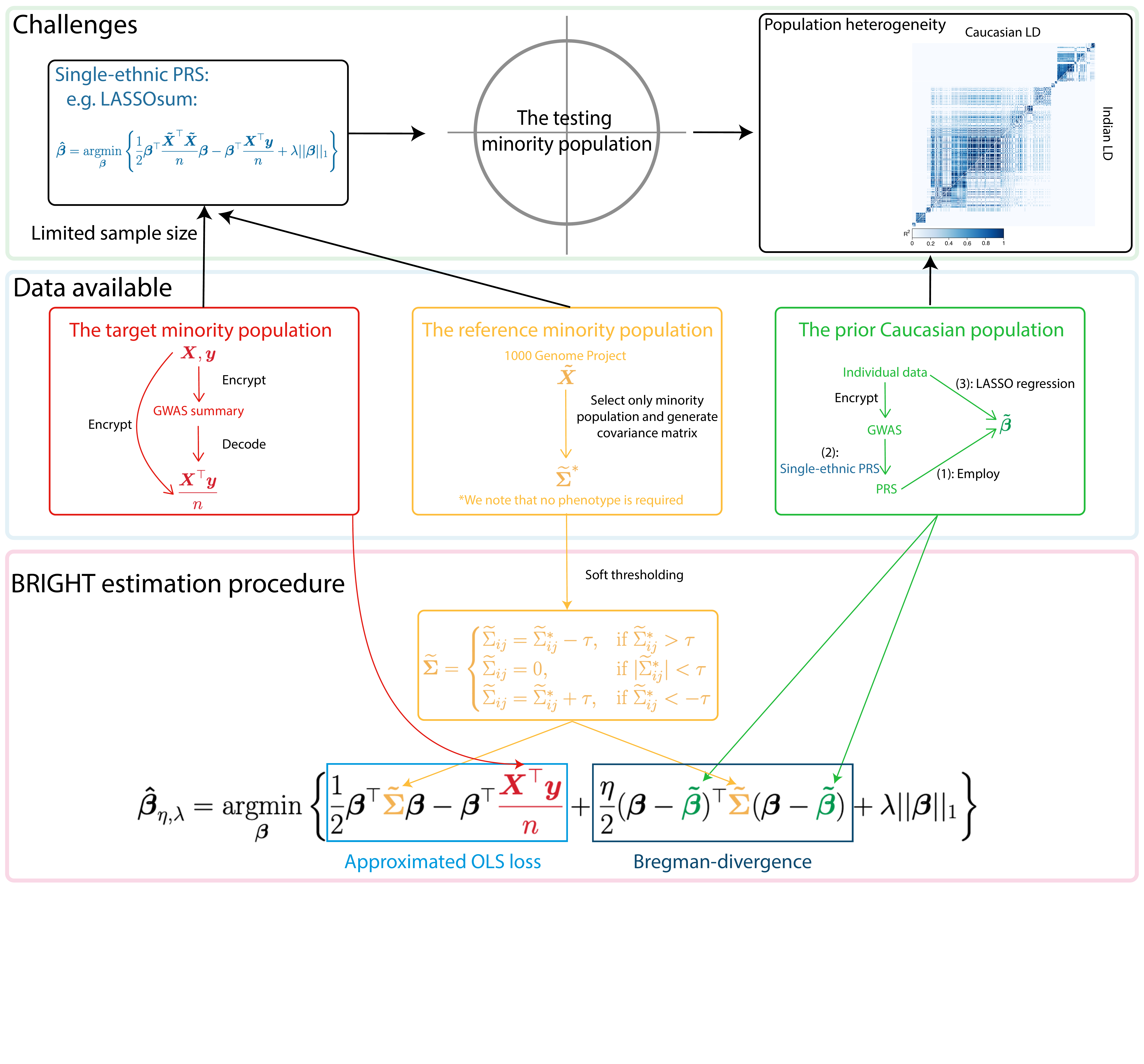}
    \caption{Graphical representation of the BRIGHT estimation procedure and the challenges of existing single-ethnic PRS methods. Summary level training data and their sources are presented in the light blue box: The encrypted target minority population summary data can be generated directly from individual-level data or through GWAS (red box); The reference minority population genotype data (no restrictions on phenotypes) can be downloaded from publicly available databases (yellow box); The prior information from Caucasian can be generated through PRS, GWAS or individual-level data (green box). The challenges of single-ethnic PRS methods based solely on target minority data or Caucasian data (light green box) are limited sample size and population heterogeneity, respectively. The proposed BRIGHT estimation procedure (light pink box) soft threshold the estimated covariance matrix; then, utilize Bregman-divergence to integrate the summary level data across different ethnic populations}
    \label{fig:diagram}
\end{figure}

A key ingredient of the proposed method, termed as
BRegman divergence-based Integrational Genetic Hazard Trans-ethnic (BRIGHT), is a metric  measuring the discrepancies between populations, which needs to (1) be based on summary statistics only; (2) incorporate minority genotype structure; (3) be computationally efficient to allow ultra-high dimensional PRS construction. To achieve these goals, we propose using a  Bregman-divergence. To balance the information, we aim to minimize the OLS of the minority data, while at the same time keeping the Bregman divergence small. This can be achieved by minimizing the penalized objective function:
\begin{align}
Q_{\eta,\lambda}(\boldsymbol\beta)&=\frac{1}{2}\boldsymbol{\beta}^\top \widetilde{\boldsymbol\Sigma}\boldsymbol{\beta}-\boldsymbol{\beta}^\top\boldsymbol{r}+\eta D_{\widetilde{\boldsymbol\Sigma}}(\boldsymbol \beta,\boldsymbol{\tilde\beta})+\lambda||\boldsymbol{\beta}||_1, \label{eq: tentative objective function}
\end{align}
where $\widetilde{\boldsymbol\Sigma}$ is a sparse covariance estimate
obtained from a minority reference data (further discussed in {\it Remark 2} below); $\boldsymbol r=\boldsymbol X^\top \boldsymbol{y}/n$ is the summary statistics obtained from the minority GWAS; $D_{\widetilde{\boldsymbol\Sigma}}(\boldsymbol \beta,\boldsymbol{\tilde\beta})=\frac{1}{2}(\boldsymbol \beta-\boldsymbol{\tilde\beta})^\top\widetilde{\boldsymbol\Sigma} (\boldsymbol \beta-\boldsymbol{\tilde\beta})$ is a covariance-adjusted squared-Mahalanobis distance (a special case of Bregman divergence) measuring the distance between $\boldsymbol \beta$ and $\boldsymbol{\tilde\beta}$, where the weighted distance  depends on the covariance structures of the minority genotypes; $\eta$ is the tuning parameter weighing the relative importance of the minority data over the Caucasian information; and $\lambda$ is the tuning parameter for the $\ell_1$ penalty. A graphic representation of the problem, available data and our solution is provided in Figure \ref{fig:diagram}.

{\it Remark 1:} The first two terms in \eqref{eq: tentative objective function}, $\frac{1}{2}\boldsymbol{\beta}^\top \widetilde{\boldsymbol\Sigma}\boldsymbol{\beta}-\boldsymbol{\beta}^\top\boldsymbol{r},$ are built upon the OLS of the minority data, 
$
\ell(\boldsymbol{\beta})=\frac{1}{2n}(\boldsymbol y-\boldsymbol X\boldsymbol{\beta})^\top(\boldsymbol y-\boldsymbol X\boldsymbol{\beta})\propto\frac{1}{2n}\boldsymbol{\beta}^\top \boldsymbol X^\top\boldsymbol X\boldsymbol{\beta}-\boldsymbol{\beta}^\top\boldsymbol{r}.
$
Using Lemma 1 in Section 3, it can be shown that $\ell(\boldsymbol{\beta}) \propto \frac{1}{2}\boldsymbol{\beta}^\top \widetilde{\boldsymbol\Sigma}\boldsymbol{\beta}-\boldsymbol{\beta}^\top\boldsymbol{r}+o_P(1)$. Moreover, 
\begin{align*}
Q_{\eta,\lambda}(\boldsymbol\beta) \propto 
(1-\eta_0) D_{\widetilde{\boldsymbol\Sigma}}(\boldsymbol \beta,\boldsymbol{\hat\beta}) + \eta_0  D_{\widetilde{\boldsymbol\Sigma}}(\boldsymbol \beta,\boldsymbol{\tilde\beta})+\frac{\lambda}{1+\eta}||\boldsymbol{\beta}||_1+o_P(1),
\end{align*}
where $\eta_0 = \eta/(1+\eta)$ and $\boldsymbol{\hat\beta}=(\boldsymbol X^\top \boldsymbol{X})^{-1}\boldsymbol X^\top \boldsymbol{y}$ is the OLS estimator on the minority data. Thus,  the proposed BRIGHT procedure provides an optimal estimate that minimizes the convex sum of the Bregman divergence between the working model and two extremes: one in which no prior information is used and the other in which no internal data is used.

{\it Remark 2:}
In our motivating settings, due to privacy considerations, researchers usually do not have access to individual genotype information, precluding the direct estimation of genotype covariance structure, $\Sigma$, using the target minority data. Our proposed solution is motivated by a commonly used strategy \citep{bush2012chapter, mak2017polygenic} in genetics. Specifically,  consider $\widetilde{\boldsymbol\Sigma}^*$, the LD estimate  (i.e. sample estimate of $\boldsymbol\Sigma$), obtained from publicly available external reference data (e.g. the subgroup from the 1000 Genome Project \citep{10002015global} with the same ethnicity as the target minority population).
For a constant $\tau > 0$, $\widetilde{\boldsymbol\Sigma}$ is defined as the soft-thresholded matrix of $\widetilde{\boldsymbol\Sigma}^*$ such that
$\widetilde{\boldsymbol\Sigma}_{ij} = sign(\widetilde{\boldsymbol\Sigma}^*_{ij})
(|\widetilde{\boldsymbol\Sigma}^*_{ij}|-\tau) 1\{|\widetilde{\boldsymbol\Sigma}^*_{ij}| \geq \tau\}$, where $1\leq i,j\leq p$, and $\widetilde{\boldsymbol\Sigma}^*_{ij}$ and $\widetilde{\boldsymbol\Sigma}^*_{ij}$ are the $i^{th}$ row and $j^{th}$ column element of the corresponding matrices.
It is worth noting that 
the dimension of the design matrices, $p$, is at the million level in genomics studies and is usually much larger than the sample size, bringing challenges in both covariance estimation \citep{wen2010using,pourahmadi2013high} and computation; the high dimensionality of $p$ dramatically increases the space as well as the time complexity of the estimation procedure. These estimation and computation inefficiencies motivate us to use a thresholding method for estimating the covariance matrix and achieving sparse structures. 

{\it Remark 3:} 
When $\eta=0$, the BRIGHT model simplifies to the LASSO regression on the target data; when $\eta=\infty$ the BRIGHT model simplifies to the prior model generated by the Caucasian population. 
The proposed method can be applied to multiple external models, identifying the most compatible external ones and diminishing less relevant ones.

{\it Remark 4:} 
The penalized objective function can be rewritten as:
\begin{align}
    Q_{\eta,\lambda}(\boldsymbol\beta)&\propto \frac{1+\eta}{2}\boldsymbol{\beta}^\top \widetilde{\boldsymbol\Sigma}\boldsymbol{\beta}-\boldsymbol{\beta}^\top(\boldsymbol{r}+\eta  \widetilde{\boldsymbol\Sigma}\boldsymbol{\tilde\beta}) +\lambda||\boldsymbol{\beta}||_1, \label{eq: Final objective function}
\end{align}
which can be optimized similarly as the classic OLS LASSO. Hence, the proposed method is computationally efficient, which is convenient for ultra-high dimensional PRS construction. Implementation details for the proposed method are provided in Supplementary Materials. 

\section{Theoretical properties: Oracle inequality} \label{Theoretical properties}

In this section, we derive high-dimensional theoretical properties of the BRIGHT estimation procedure, where the covariate dimension is allowed to grow at a significantly faster rate than the sample sizes, $p\gg n$, while both are allowed to grow to infinity. Since the main objective of PRS methods is risk prediction,  we investigate the oracle inequality for the BRIGHT oracle rate under the random design squared error. Compared to the fixed design setting of oracle inequality proofs \citep{bunea2007sparsity,jiang2016variable}, where the model prediction consistency is evaluated on the training data, random design evaluates the model on independent testing data, and takes expectation over outcome and predictors. The excess risk or prediction error under the random design for a linear model is defined as
$$\mathcal{E}(\boldsymbol{\hat\beta}_{\eta,\lambda})=E_{\boldsymbol\Psi, y}\{(\boldsymbol\Psi^\top\boldsymbol{\hat\beta}_{\eta,\lambda}-y)^2\}=(\boldsymbol{\hat\beta}_{\eta,\lambda}-\boldsymbol\beta_0)^\top\boldsymbol\Sigma(\boldsymbol{\hat\beta}_{\eta,\lambda}-\boldsymbol\beta_0),$$
where $\boldsymbol\Psi$ is the random vector generating the target minority genotype matrix, $\boldsymbol X$, and $\boldsymbol{\hat\beta}_{\eta,\lambda}$ is the optimization result for the BRIGHT objective function, $\boldsymbol{\hat\beta}_{\eta,\lambda}=\argmax_{\boldsymbol\beta}Q_{\eta,\lambda}(\boldsymbol\beta)$.
In the following proof, we establish a refined upper bound for the BRIGHT excess risk, which holds with probability tending to 1 as $n$ goes to infinity. Finally, we will discuss the new oracle rate compared to the well-known oracle rate $r_{LASSO}=|\mathcal{S}_0|\log p/n$ \citep{van2008high,bickel2009simultaneous}. Without loss of generality, we assume $E(\boldsymbol\Psi)=\boldsymbol 0$ and $\boldsymbol X$  are centered and standardized. 

To derive estimation consistency of $\widetilde{\boldsymbol\Sigma}$, we assume the following three conditions:

{\it \textbf{Condition 1} (Bounded predictor condition):
 $\sup_{j} \Psi_j<K$ almost sure for some scalar $K$, where $\Psi_j$ is the $j^{th}$ element of random vector $\boldsymbol \Psi$.
} 

{\it \textbf{Condition 2} (Sparse predictor covariance condition):
$\Sigma_{ii}\leq K_2$ for all $1\leq i\leq p$ and some constant $K_2$; $\max_i\sum_{j=1}^p |\Sigma_{ij}|^q\leq c_0$ for all $0\leq q<1$, and some constant $c_0$ indicating the sparsity. 
}

{\it \textbf{Condition 3} (High dimensional condition):
${\log p}/{n}=o(1)$.
}

Condition 1 giving a uniform bound for the predictors is originally assumed in \cite{van2008high} for the proof of random design LASSO excess risk. In genetics, the elements in the genotype matrix are bounded; therefore Condition 1 is automatically satisfied.
Condition 2, originally employed by \cite{bickel2008covariance}, is used to control the sparsity of the underlying true predictor covariance matrix; $\boldsymbol\Sigma$ satisfying Condition 2 is a class of ``approximately sparse'' covariance matrices with $q=0$ corresponding to a class of truly sparse matrices \citep{rothman2009generalized}. In genetics, the correlation between predictors depends on their physical distance on the genome, and the correlation decays to zero as the physical distance increases. Thus, even the stronger truly sparse matrix assumption with $q=0$ is satisfied. 
Condition 3 allows the diverging speed of $p$ to be as fast as $\exp(n)$.  

Following Theorem 1 of \cite{rothman2009generalized}, {\it \textbf{Lemma 1}} below summarizes the property of the soft thresholded covariance matrix:

{\it \textbf{Lemma 1.}
Suppose $\widetilde{\boldsymbol\Sigma}$ is generated from soft thresholding with hyperparameter $\tau$, and Conditions 1-3 are met. Then for sufficiently large $M'$, if $\tau=M'\sqrt{\frac{\log p}{\tilde n}}$, we have
$$\max_{i,j}|\widetilde\Sigma_{ij}-\Sigma_{ij}|=O_P\left( c_0(\frac{\log p}{\tilde n})^{(1-q)/2}\right),$$
where $\tilde n$ is the sample size of the external reference minority data.
} 


In the rest of this section, we will discuss the main theorem, for which the following four additional conditions are imposed:

{\it \textbf{Condition 4} ($\Sigma$ compatibility condition):
 There exists $\phi_0>0$ and for all $\boldsymbol\delta$ satisfying $||\boldsymbol\delta^{\mathcal{S}_0^c}||_1\leq3||\boldsymbol\delta^{\mathcal{S}_0}||_1$, it holds that
 $$||\boldsymbol{\delta}^{ \mathcal{S}_0}||_1^2\leq(\boldsymbol{\delta}^\top\boldsymbol \Sigma\boldsymbol{\delta})|\mathcal{S}_0|/\phi_0^2,$$
 where vectors with $\mathcal{S}_0$ or $\mathcal{S}_0^c$ on the top right mean the vectors only consist of elements with indexes within the corresponding index set; $|\mathcal{S}_0|$ represents the cardinality of $\mathcal{S}_0$.} 

{\it \textbf{Condition 5} (Sub-exponential error condition):
 There exists a constant $M>0$ such that $\sup_{i}E(|\epsilon_i|^m)\leq m!M^m$ for $m=2,3,\dots$} 

{\it \textbf{Condition 6} (Convergent prior condition):
$||\boldsymbol{\tilde\beta}-\boldsymbol\beta_0||_{\infty}=O_P( \sqrt{\frac{\log p}{n_c}})$, where $n_c$ is the sample size of the Caucasian population that generated $\boldsymbol{\tilde\beta}$.} 

{\it \textbf{Condition 7} (Sparse coefficient condition):
$|\mathcal{S}_0|/\phi_0^2= o\{(\frac{n}{\log p})^{1/2}\}$ and $||\boldsymbol{\beta}_0||_1^2=o\{(\frac{n}{\log p})^{1/2}\}$.} 

Condition 4 is used to restrict the predictors from being extremely highly correlated. This condition employed in \cite{buhlmann2011statistics} is also related to the restricted eigenvalue condition used in \cite{bickel2009simultaneous}, \cite{koltchinskii2009dantzig} and  \cite{jiang2016variable}. We note that the difference between the compatibility condition and the restricted eigenvalue condition is that: (1) the former condition restricts the underlying true covariance structure, while the latter condition restricts the estimated covariance structure; and (2) the former condition is strictly weaker than the latter condition. For more discussions on the compatibility and restricted eigenvalue conditions, please refer to \cite{van2009conditions}.
Condition 5 assumes the sub-exponential distribution of the error term, which controls its tail probability, $Pr\{|\epsilon| \geq t\} \leq 2 \exp \left(-\frac{t^2}{2(M^2+Mt)}\right) \text { for all } t > 0$ (i.e. Bernstein inequality). We note that this condition is strictly weaker than the normal or sub-Gaussian noise assumption for the error term.
Condition 6 controls the quality of the prior information, $\boldsymbol{\tilde\beta}$. It is noteworthy that this condition is automatically satisfied, if the $\boldsymbol{\tilde\beta}$ is generated from the LASSO regression model on individual-level Caucasian data and if the true effect sizes for the two populations are assumed to be homogeneous \citep{buhlmann2011statistics}.
The first part of Condition 7 controls the true effect to be mostly zero, the second part of Condition 7 provides a $\ell_1$ norm bound for the true effect sizes. We note that similar conditions are assumed in \cite{van2008high} and \cite{buhlmann2011statistics}.

Then, we present the main theorem as follows; the detailed proof can be found in the Supplementary Materials.

{\it \textbf{Theorem 1}
When all above seven conditions are satisfied and let $$\lambda=A\left[\sqrt{\frac{logp}{n}}+\eta\sqrt{\frac{\log p}{n_c}}\left\{c_0(\frac{\log p}{\tilde n})^{(1-q)/2}+c_0\right\}\right]$$ for some large scalar $A$, $\tilde \lambda=(\frac{\log p}{n})^{1/2}$ and $\tilde \lambda'=c_0(\frac{\log p}{\tilde n})^{(1-q)/2}$. Then, with the probability at least $1-2\exp{(-Clogp)}$, the BRIGHT estimator $\boldsymbol{\hat\beta}_{\eta,\lambda}$ satisfies:
$$ \mathcal{E}(\boldsymbol{\hat\beta}_{\eta,\lambda})\leq\frac{36A^2|\mathcal{S}_0|}{\phi_0^2B}\left[\frac{1}{(1+\eta)^2}\frac{{logp}}{{n}}+\frac{\eta^2}{(1+\eta)^2}\frac{{\log p}}{{n_c}}\left\{c_0(\frac{\log p}{\tilde n})^{(1-q)/2}+c_0\right\}^2\right]+\frac{6(\tilde\lambda+\tilde\lambda')||\boldsymbol{\beta}_0||_1^2}{B(1+\eta)},$$
where $B=1-\{96\tilde\lambda|\mathcal{S}_0|/\phi_0^2+48(1+\eta)\tilde\lambda'|\mathcal{S}_0|/\phi_0^2\}/(1+\eta)$ converges to 1 as $n\rightarrow\infty$ and $C$ converges to some positive constant as $n\rightarrow\infty$ (for detailed formulation please refer to the proof in Supplementary Materials).
}

Theorem 1 provides a general oracle inequality for the excess risk $\mathcal{E}(\boldsymbol{\hat\beta}_{\eta,\lambda})$. When $n,n_c,\tilde n, p \rightarrow\infty$, the excess risk converges to zero with probability tending to one. The first term in the oracle rate is similar to what has been established for lasso regression, $|\mathcal{S}_0|\frac{\log p}{n}$;  When no information is borrowed from the Caucasian population, $\eta=0$, the first term above reduces to the classic lasso rate. However, when additional information is borrowed from the Caucasian population, the first term of BRIGHT oracle rate lies somewhere between $|\mathcal{S}_0|\frac{\log p}{n}$ and $|\mathcal{S}_0|\frac{\log p}{n_c}$ for some $\eta$, and with a careful selection of $\eta$ it can achieve a better rate smaller than either of the above two. We note that the $\frac{\log p}{n_c}$ directly came from Condition 6, for milder prior convergence conditions the above theorem still holds with a replacement of $\frac{\log p}{n_c}$. The second term in the above bound is the price to pay for using $\widetilde{\boldsymbol\Sigma}$ to approximate the original $\boldsymbol{\hat\Sigma}=\frac{\boldsymbol X^\top \boldsymbol X}{n}$; we note that as $n\rightarrow\infty$ the second term will still converge to zero but the rate highly depends on the diverging speed of $||\boldsymbol{\beta}_0||_1^2$ in the assumption.

\section{Simulation}\label{simulation}

We compare the prediction and variable selection performance of the BRIGHT estimation procedure with three competing methods: a trans-ethnic PRS method, PRS-CSx \citep{ruan2021improving}; a  single-ethnic prediction model, LASSOsum; and a null Principal Component (PC) model.

We simulate phenotypes using the haplotypes from a psoriasis genetic cohort with one Caucasian and two independent South Asian datasets to preserve the real predictor structures. We consider the South Asians as the minority target population and we treat one of the South Asian datasets as training and the other as testing. 
We utilize the first 10,000 SNPs from chromosome 1 and chromosome 2, a total of $p=20,000$ predictors, and denote the genotype matrices for Caucasian, South Asian training and South Asian testing data as $\boldsymbol X_c$, $\boldsymbol X$, and $\boldsymbol X_t$ with sample sizes 11,675, 1,817 and 2516, respectively.
We simulate outcomes by $\boldsymbol{y}\sim N(\boldsymbol X\boldsymbol \beta_0, \sigma^2)$, $\boldsymbol{y_t}\sim N(\boldsymbol X_t\boldsymbol \beta_0, \sigma^2)$ and $\boldsymbol{y}_c\sim N(\boldsymbol X_c\boldsymbol \beta_c, \sigma_c^2)$, where $\boldsymbol \beta_0$ and $\boldsymbol \beta_c$ are the effects of the South Asian and Caucasian populations;  $\sigma=\widehat{Var}(\boldsymbol X\boldsymbol \beta_0)(1/h^2-1)$ and $\sigma_c=\widehat{Var}(\boldsymbol X_c\boldsymbol \beta_c)(1/h^2-1)$, with $h^2={{Var}(\boldsymbol X\boldsymbol \beta_0)}/{{Var}(\boldsymbol y)}$
controlling the heritability (signal to noise ratio). 
Plink1.9 \citep{purcell2007plink} is applied to obtain the GWAS summary information. For the Caucasian population,  LASSOsum is then implemented to obtain the PRS coefficient estimates, $\boldsymbol{\tilde\beta}$, which we treat as our prior information. The simulations are repeated 100 times.
\subsection{Prediction performance evaluation}

To evaluate the prediction performance,  we randomly sample the non-zero index set, $\mathcal{S}_0$; then, generate Caucasian and South Asian non-zero effects from a bivariate Gaussian distribution: $$\begin{pmatrix}\beta_{c}^{\mathcal{S}_0}\\\beta_{0}^{\mathcal{S}_0}\end{pmatrix}\sim N\left(\begin{bmatrix}0\\0\end{bmatrix},\begin{bmatrix}\frac{h^2}{|\mathcal{S}_0|} & \frac{\rho h^2}{|\mathcal{S}_0|}\\ \frac{\rho h^2}{|\mathcal{S}_0|} & \frac{h^2}{|\mathcal{S}_0|} \end{bmatrix}\right),$$
where $\rho$ controls the discrepancy between the Caucasian and minority effect sizes; $\rho=1$ means the effect sizes for the two populations are exactly the same and $\rho=0$ means the effect sizes are independent. The prediction performances are compared based on $R^2$, the squared correlation between the observed and predicted outcome.

We explore the effect of polygenicity by varying the number of non-zero effect sizes from 20 to 200, corresponding to low and high polygenicity scenarios; we explore different heritability, $h^2=0.7, 0.5, 0.3$, representing high, medium and low signal to noise ratio; and we also explore the different heterogeneity levels between the two population by setting $\rho=1,0.9,0.7$, representing homogeneous, low heterogeneity and high heterogeneity.

\begin{figure}[h]
    \centering
    \subfigure[$h^2=0.7$, $|\mathcal{S}_0|=20$]{\includegraphics[width=0.31\textwidth]{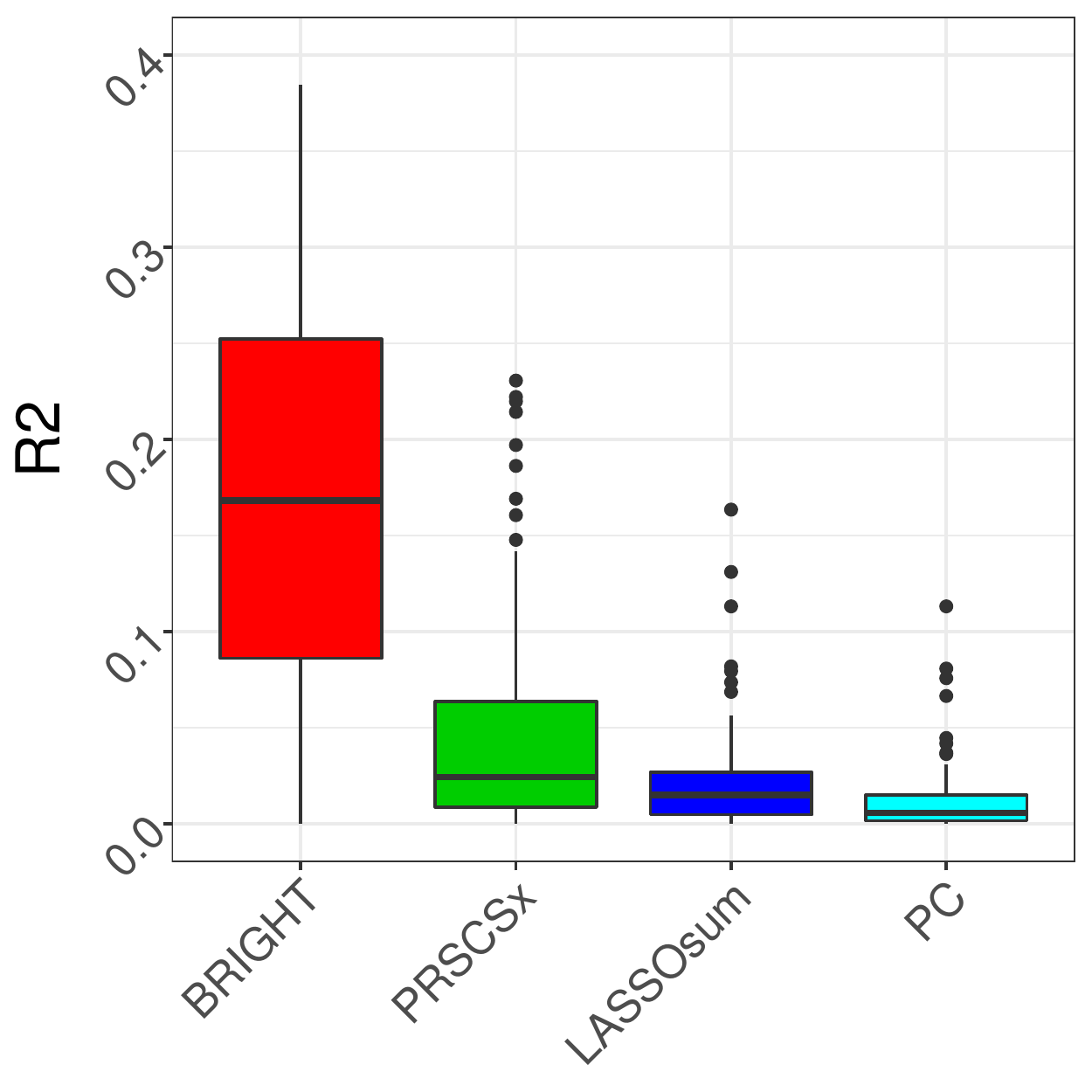}}
    \subfigure[$h^2=0.5$, $|\mathcal{S}_0|=20$]{\includegraphics[width=0.31\textwidth]{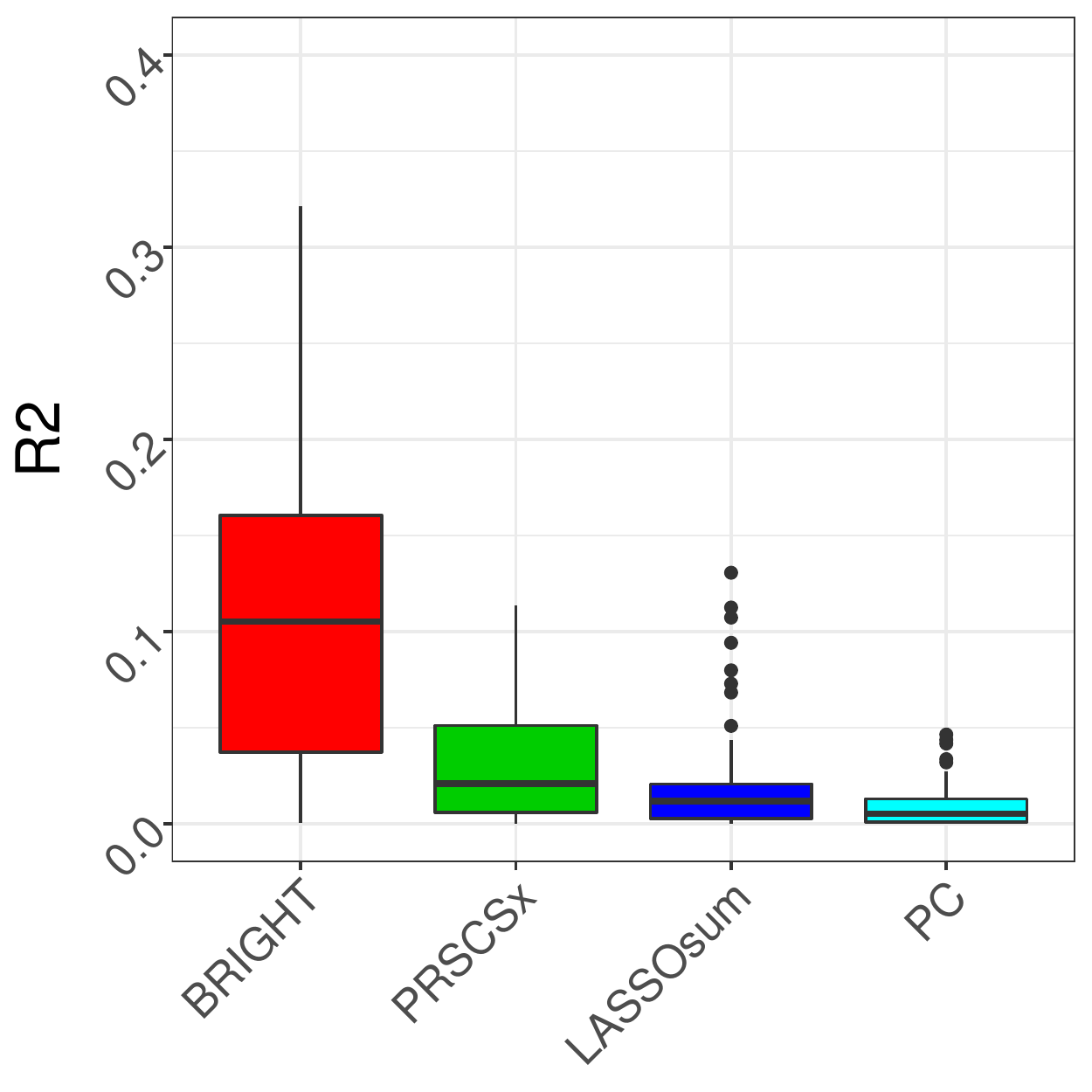}}
    \subfigure[$h^2=0.3$, $|\mathcal{S}_0|=20$]{\includegraphics[width=0.31\textwidth]{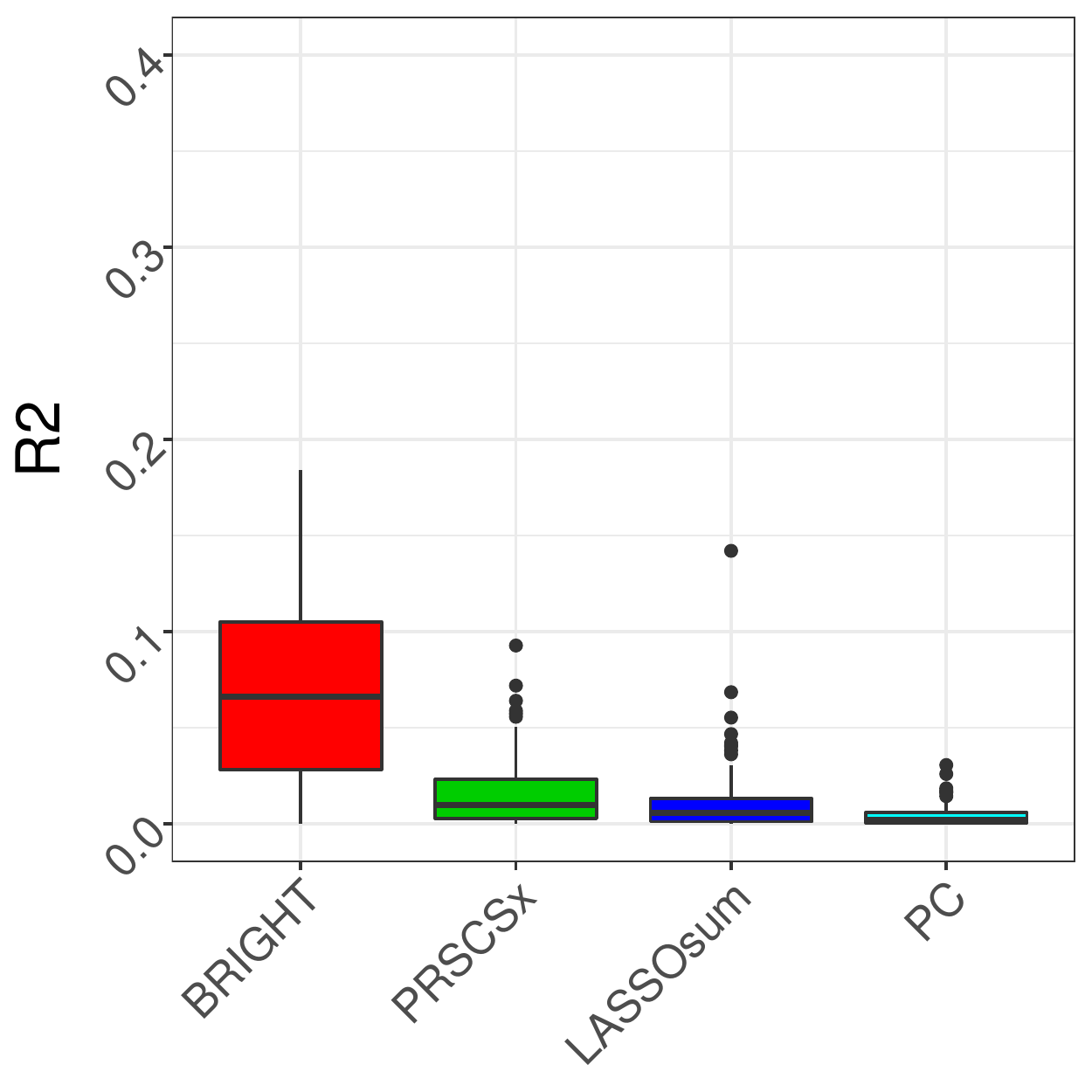}}
    \subfigure[$h^2=0.7$, $|\mathcal{S}_0|=200$]{\includegraphics[width=0.31\textwidth]{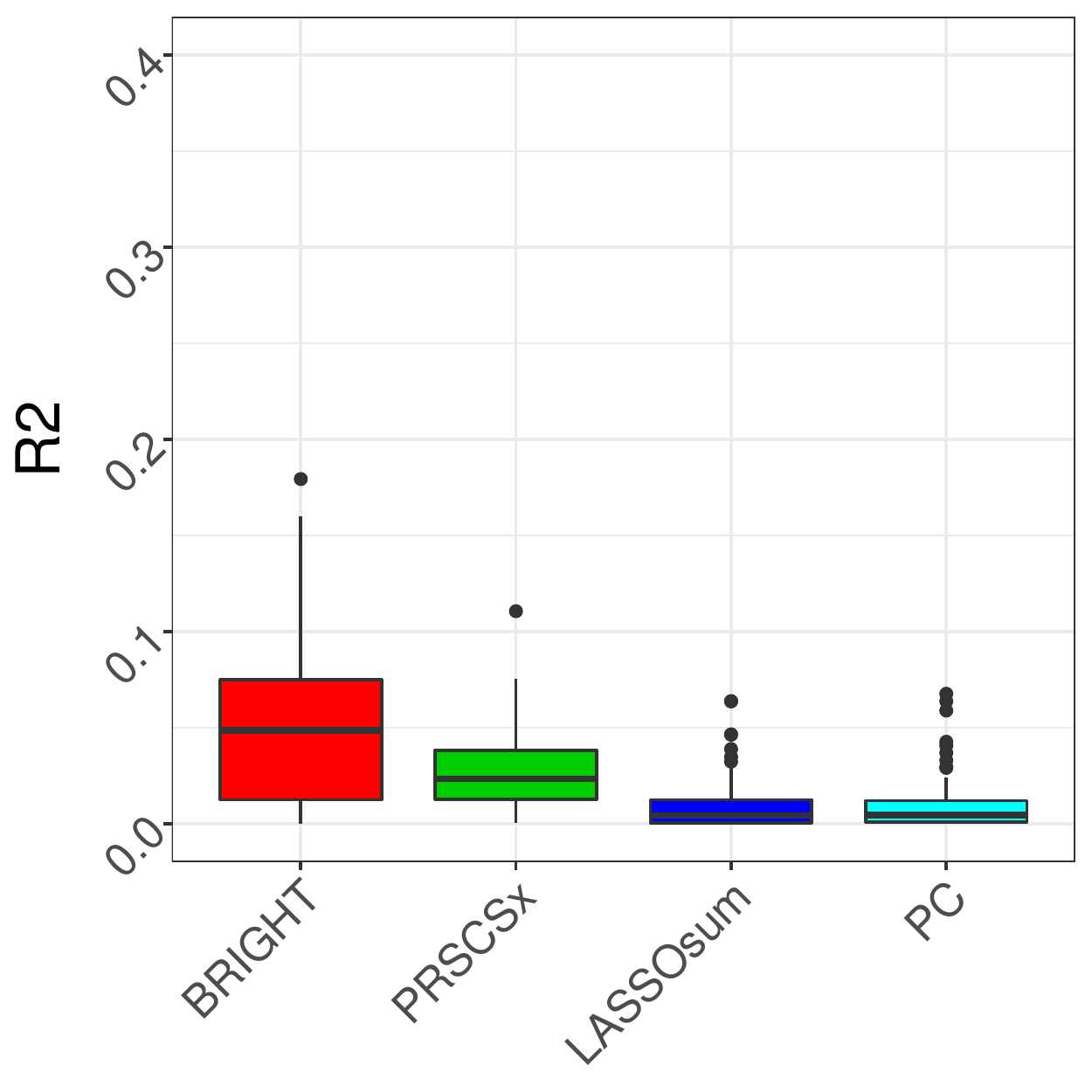}}
    \subfigure[$h^2=0.5$, $|\mathcal{S}_0|=200$]{\includegraphics[width=0.31\textwidth]{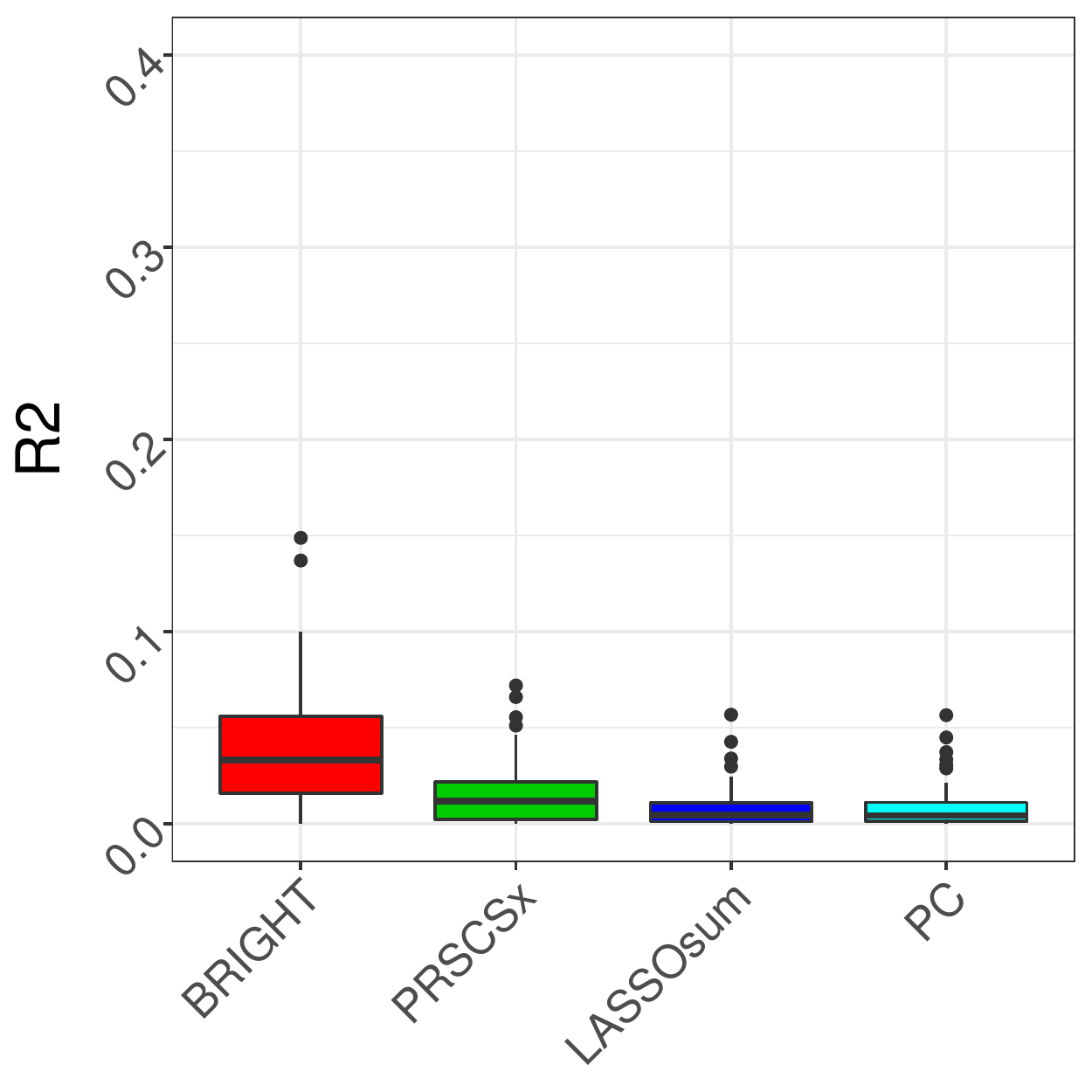}}
    \subfigure[$h^2=0.3$, $|\mathcal{S}_0|=200$]{\includegraphics[width=0.31\textwidth]{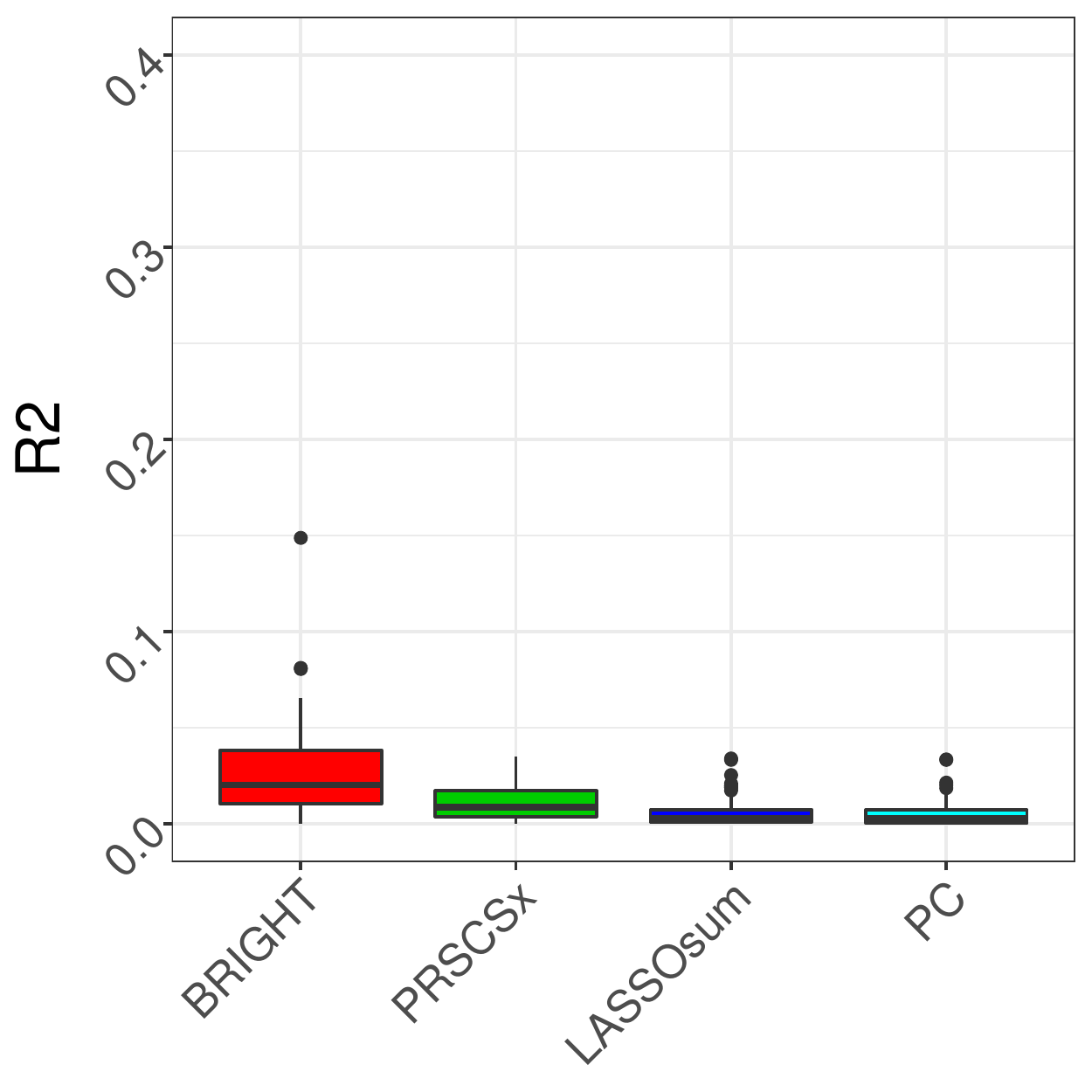}}
    \caption{Simulation results for homogeneous Caucasian and South Asian genetic effects, $\rho=1$. $h^2$ represents the heritability or signal to noise ratio and $|\mathcal{S}_0|$ represents the polygenicity level. Simulation figures from left to right are associated with a decreasing signal to noise ratio simulation setting and from up to down figures are having larger polygenicity levels.}
    \label{fig:rho=1}
\end{figure}

\begin{figure}[h]
    \centering
    \subfigure[$h^2=0.7$, $|\mathcal{S}_0|=20$]{\includegraphics[width=0.31\textwidth]{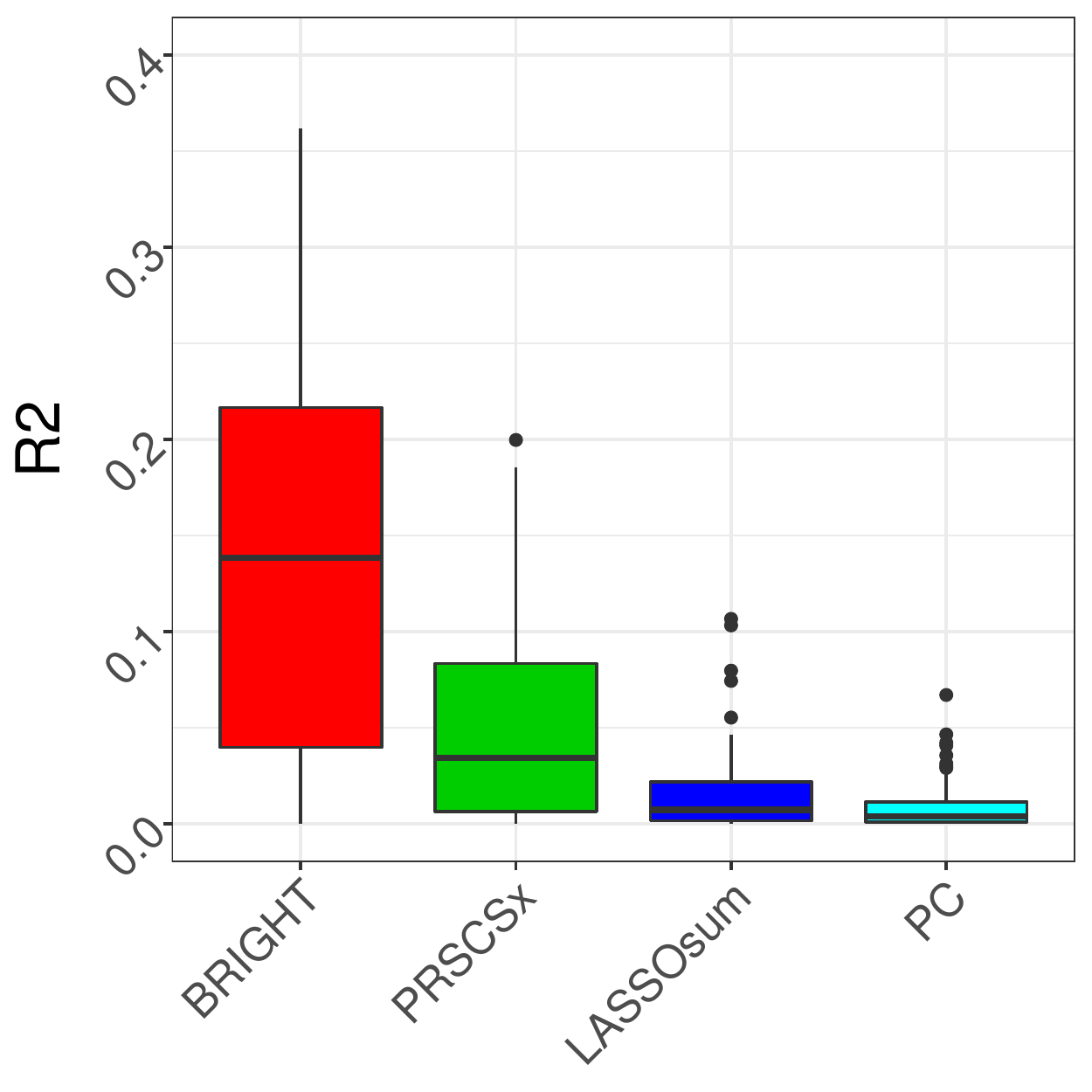}}
    \subfigure[$h^2=0.5$, $|\mathcal{S}_0|=20$]{\includegraphics[width=0.31\textwidth]{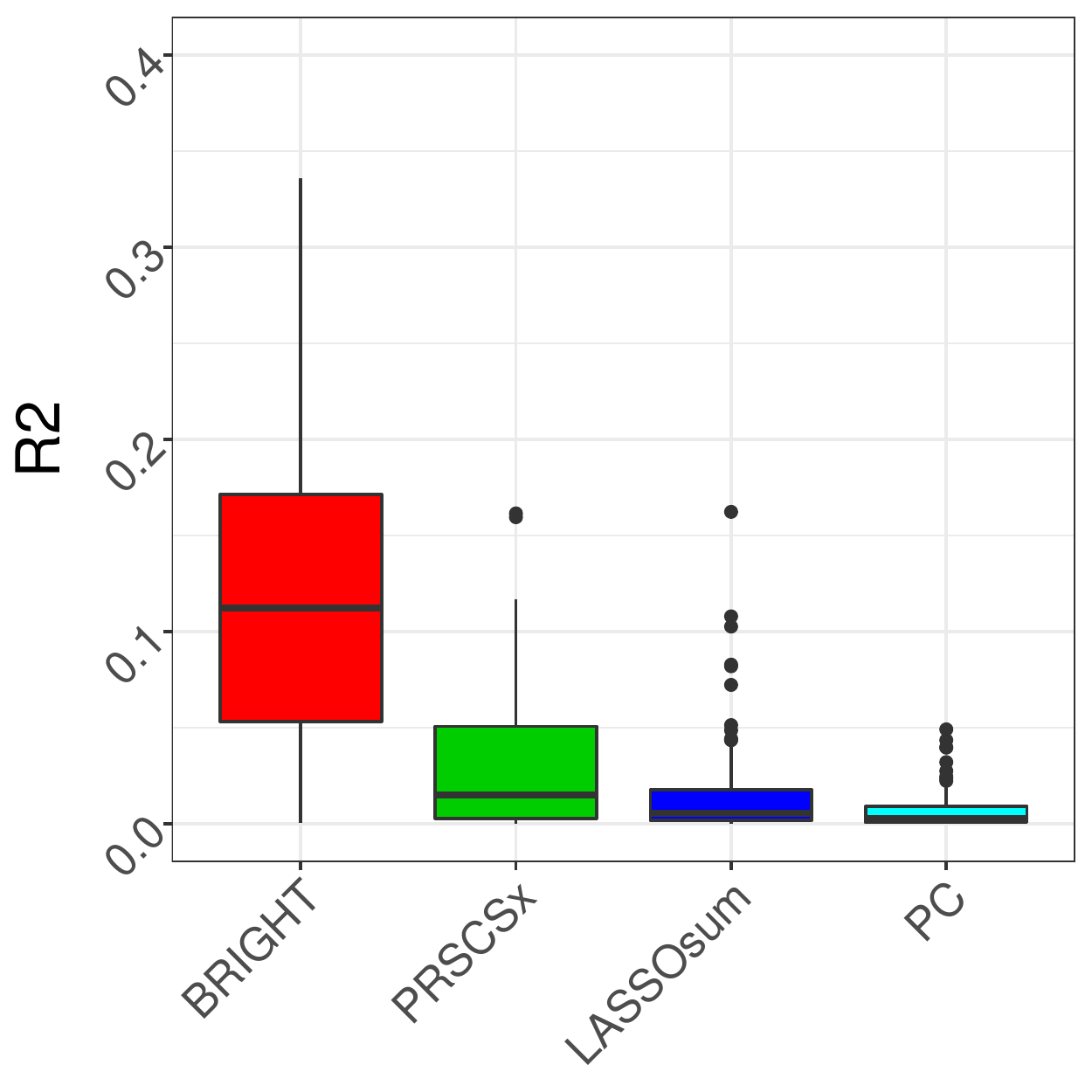}}
    \subfigure[$h^2=0.3$, $|\mathcal{S}_0|=20$]{\includegraphics[width=0.31\textwidth]{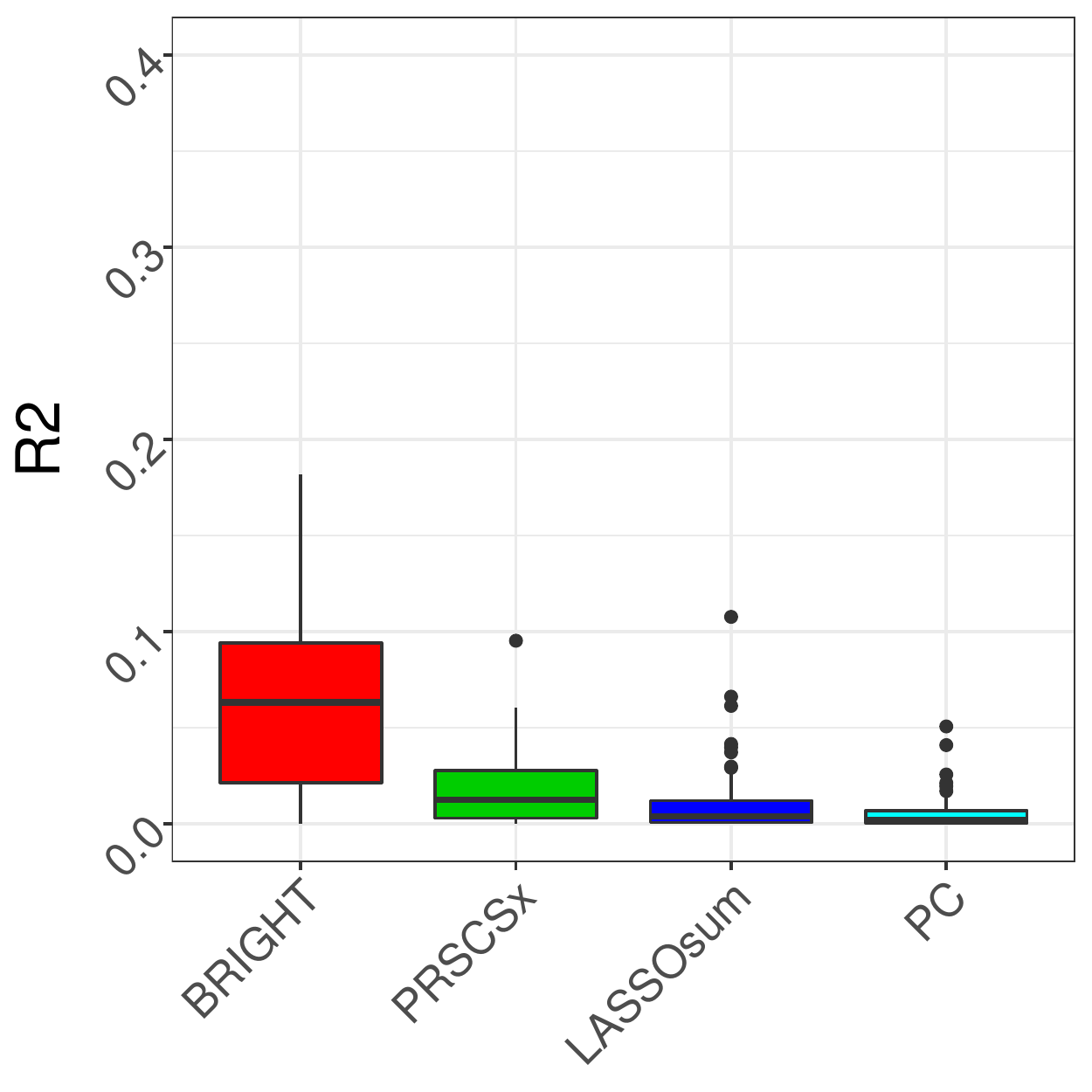}}
    \subfigure[$h^2=0.7$, $|\mathcal{S}_0|=200$]{\includegraphics[width=0.31\textwidth]{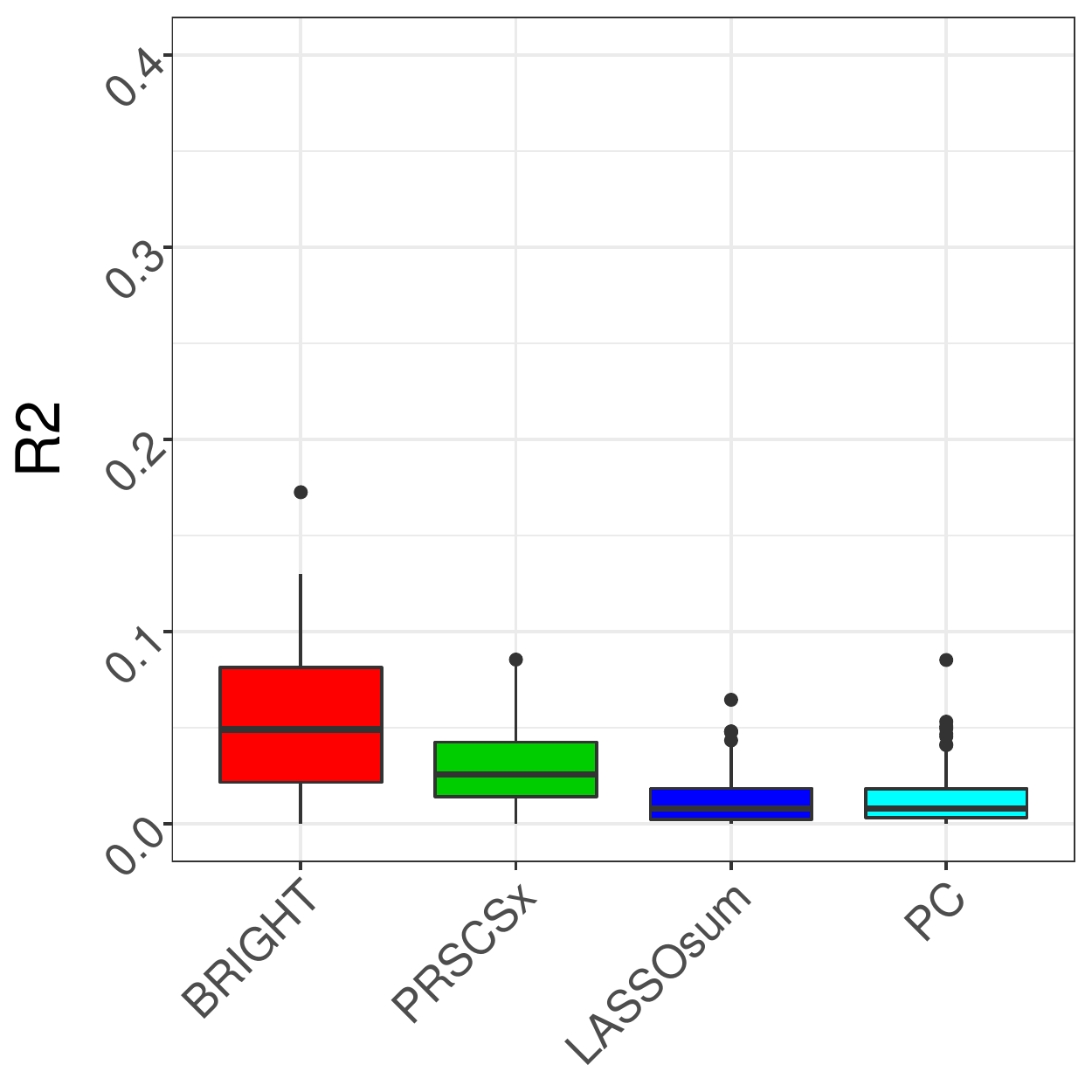}}
    \subfigure[$h^2=0.5$, $|\mathcal{S}_0|=200$]{\includegraphics[width=0.31\textwidth]{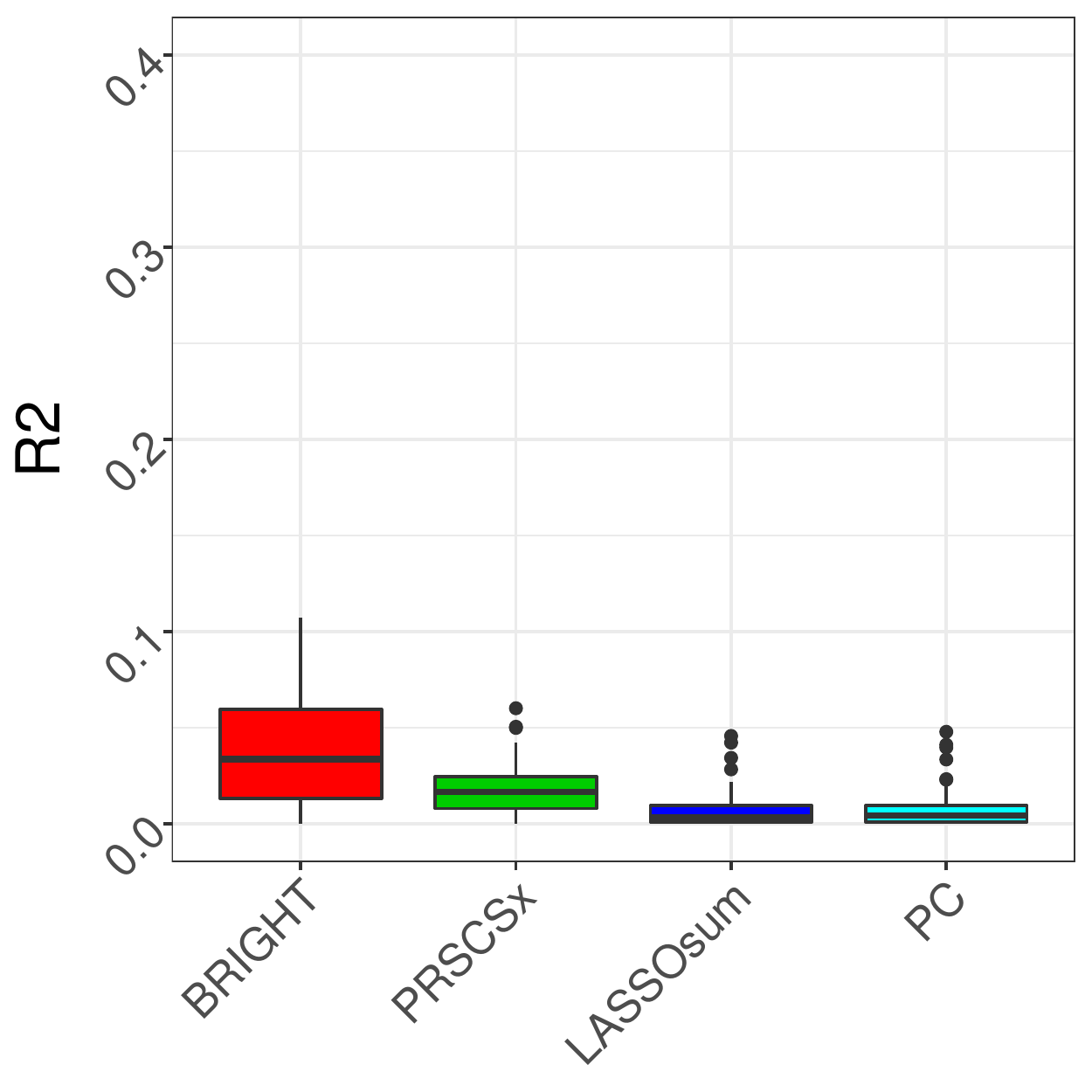}}
    \subfigure[$h^2=0.3$, $|\mathcal{S}_0|=200$]{\includegraphics[width=0.31\textwidth]{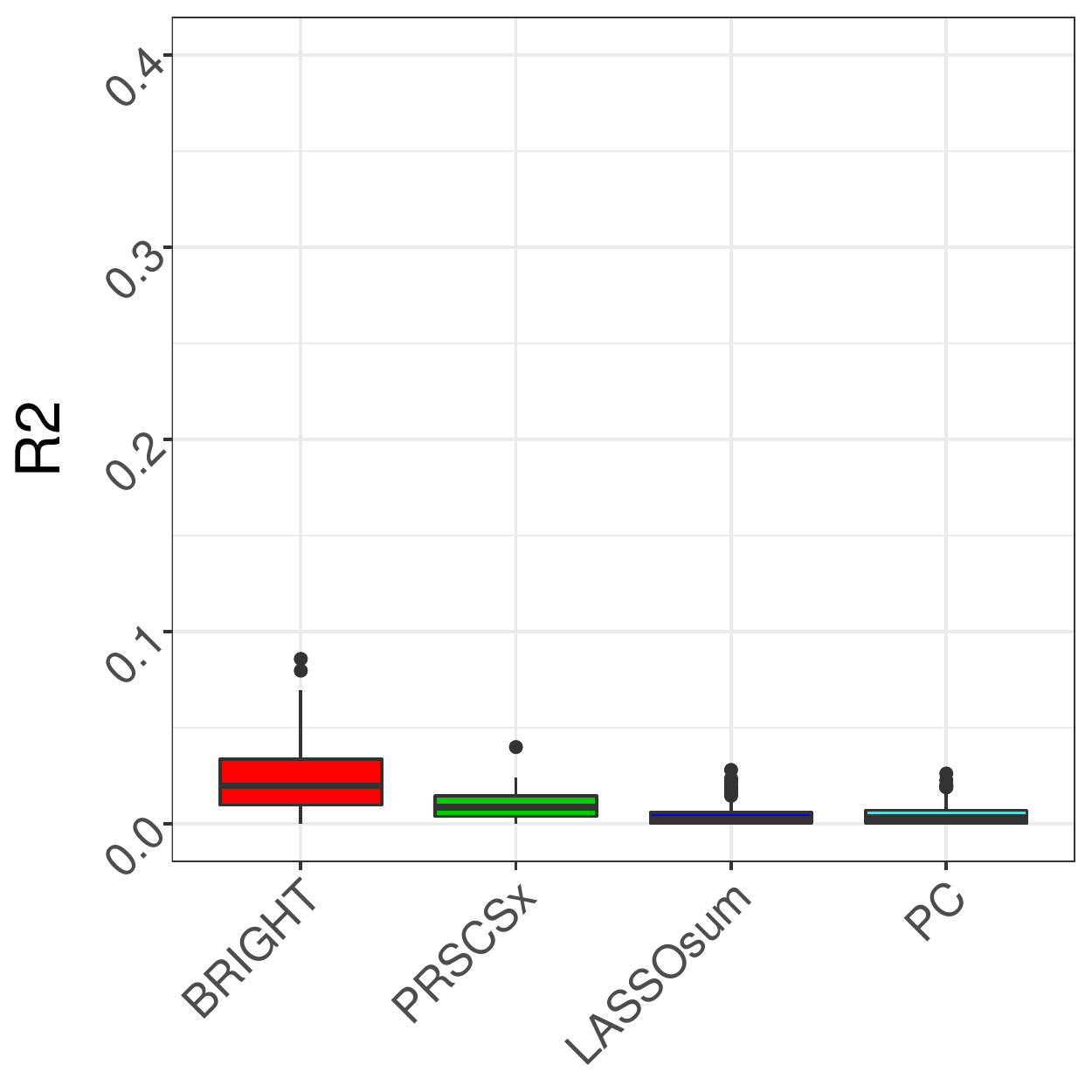}}
    \caption{Simulation results for medium heterogeneous Caucasian and South Asian genetic effects, $\rho=0.9$. $h^2$ represents the heritability or signal to noise ratio and $|\mathcal{S}_0|$ represents the polygenicity level. Simulation figures from left to right are associated with a decreasing signal to noise ratio simulation setting and from up to down figures are having larger polygenicity levels.}
    \label{fig:rho=0.9}
\end{figure}

\begin{figure}[h]
    \centering
    \subfigure[$h^2=0.7$, $|\mathcal{S}_0|=20$]{\includegraphics[width=0.31\textwidth]{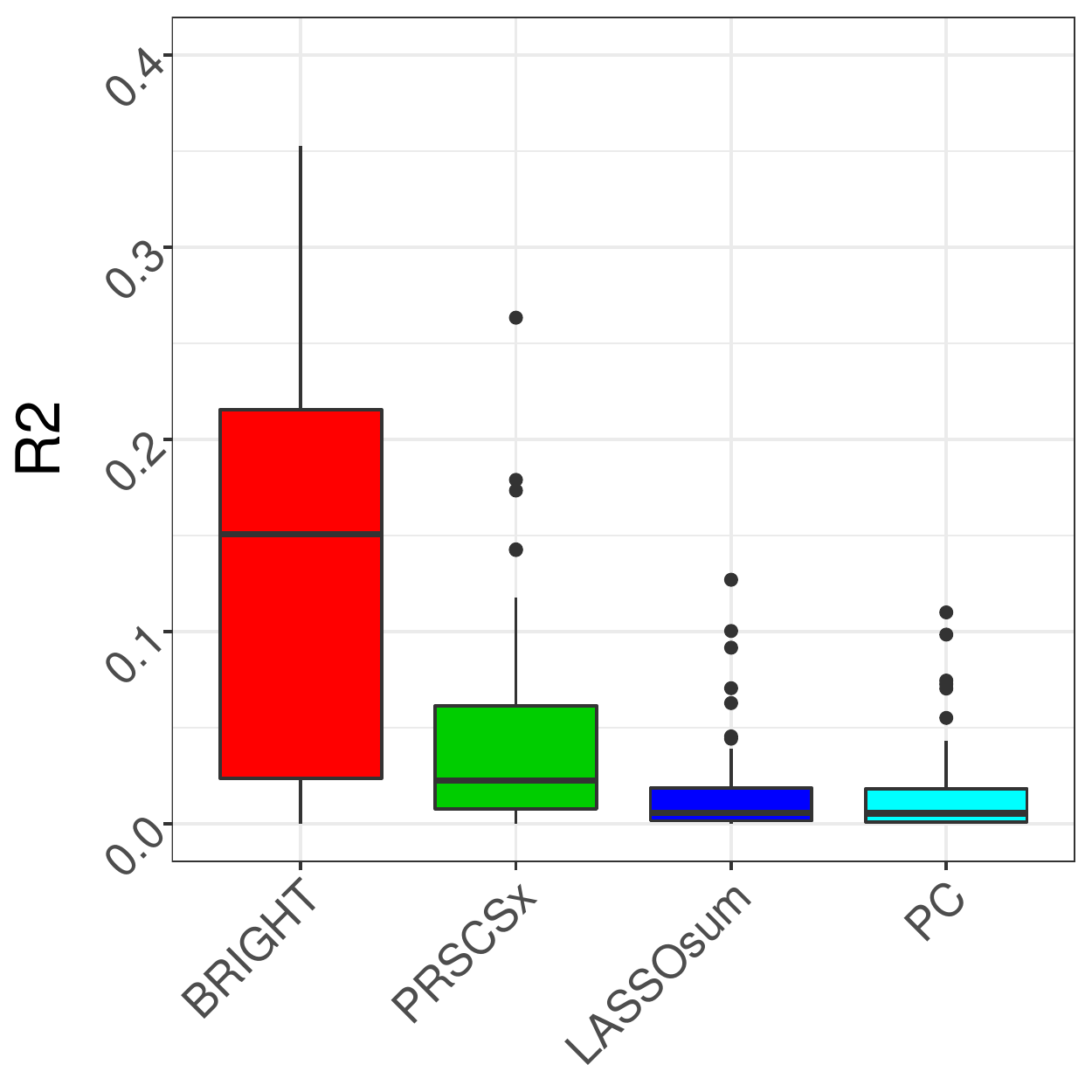}}
    \subfigure[$h^2=0.5$, $|\mathcal{S}_0|=20$]{\includegraphics[width=0.31\textwidth]{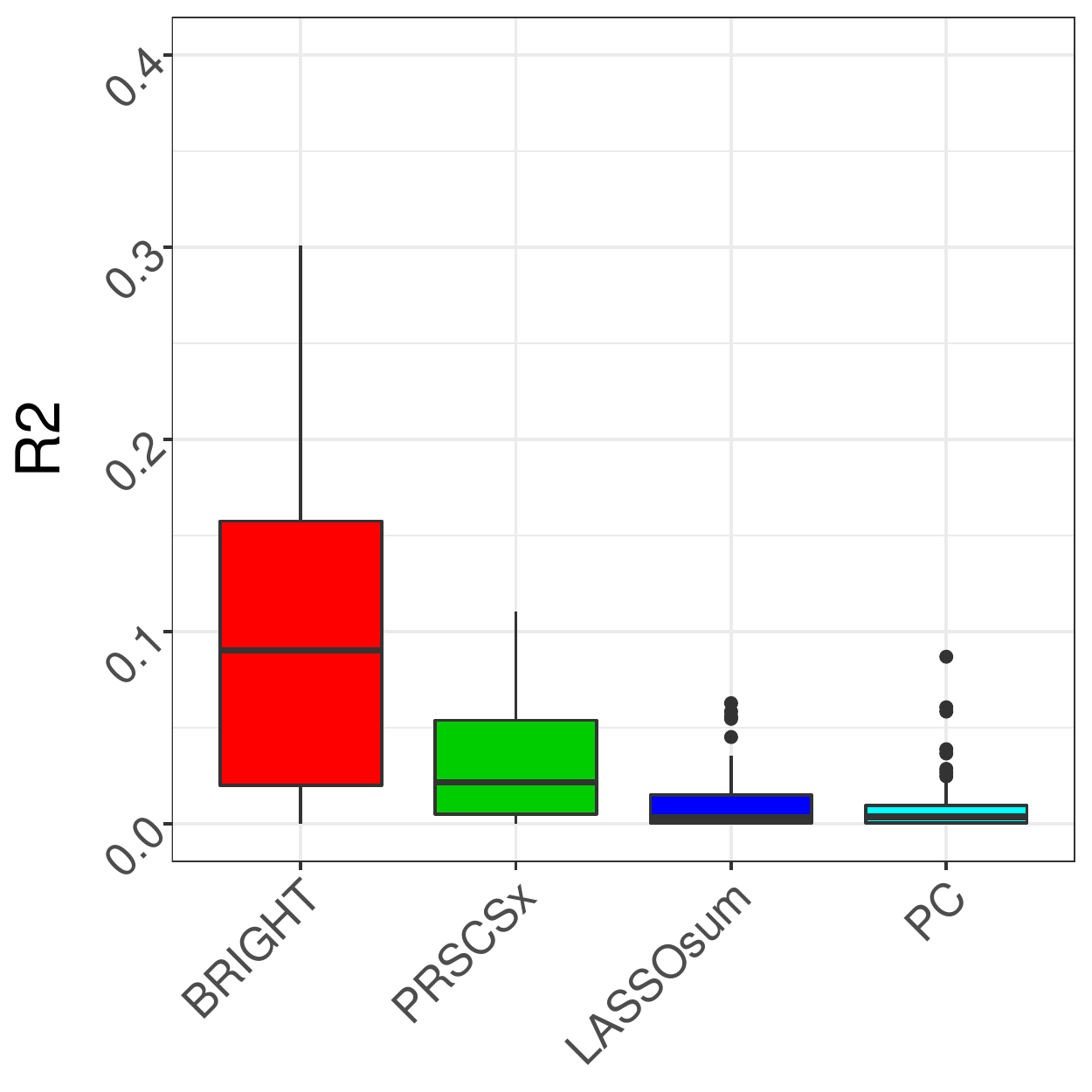}}
    \subfigure[$h^2=0.3$, $|\mathcal{S}_0|=20$]{\includegraphics[width=0.31\textwidth]{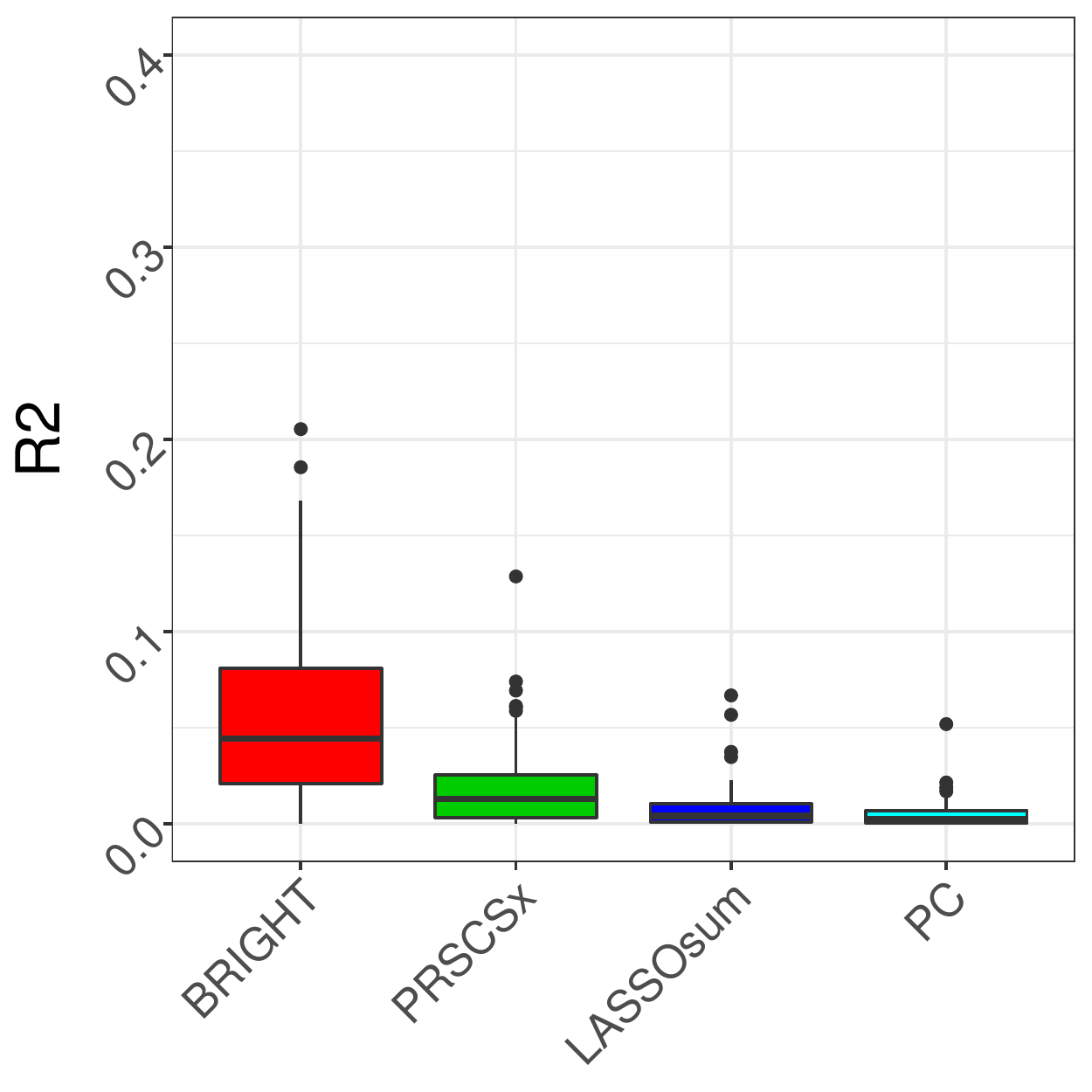}}
    \subfigure[$h^2=0.7$, $|\mathcal{S}_0|=200$]{\includegraphics[width=0.31\textwidth]{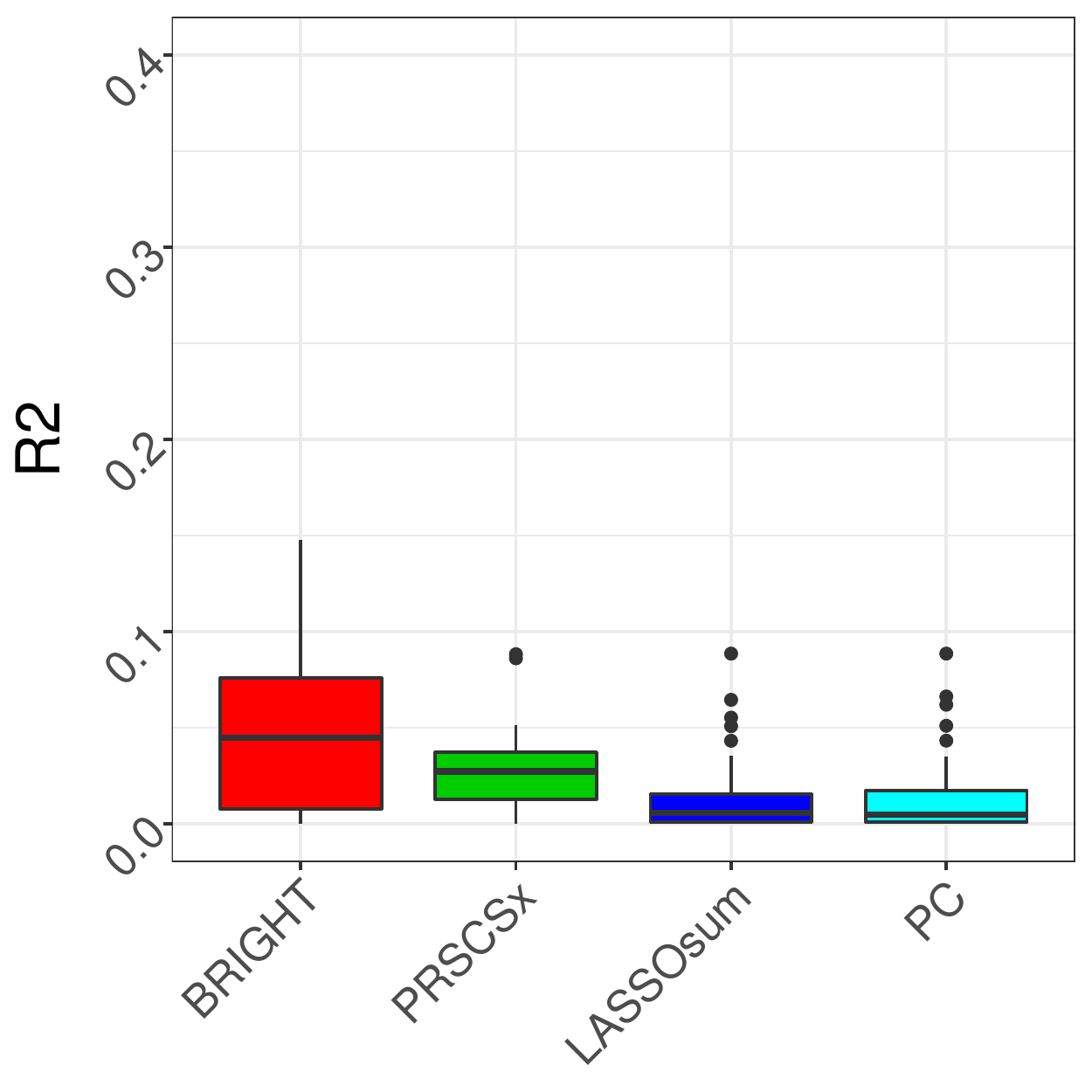}}
    \subfigure[$h^2=0.5$, $|\mathcal{S}_0|=200$]{\includegraphics[width=0.31\textwidth]{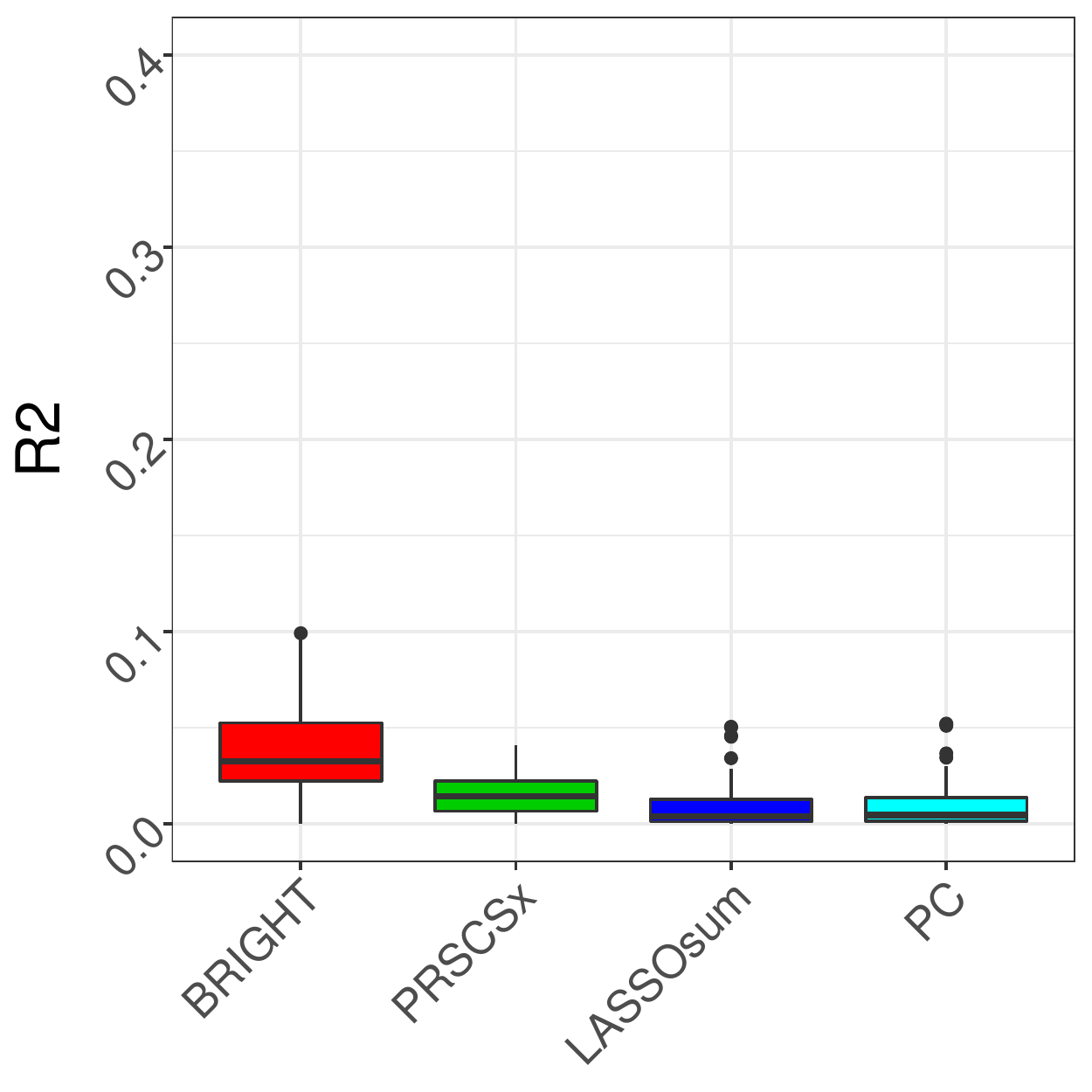}}
    \subfigure[$h^2=0.3$, $|\mathcal{S}_0|=200$]{\includegraphics[width=0.31\textwidth]{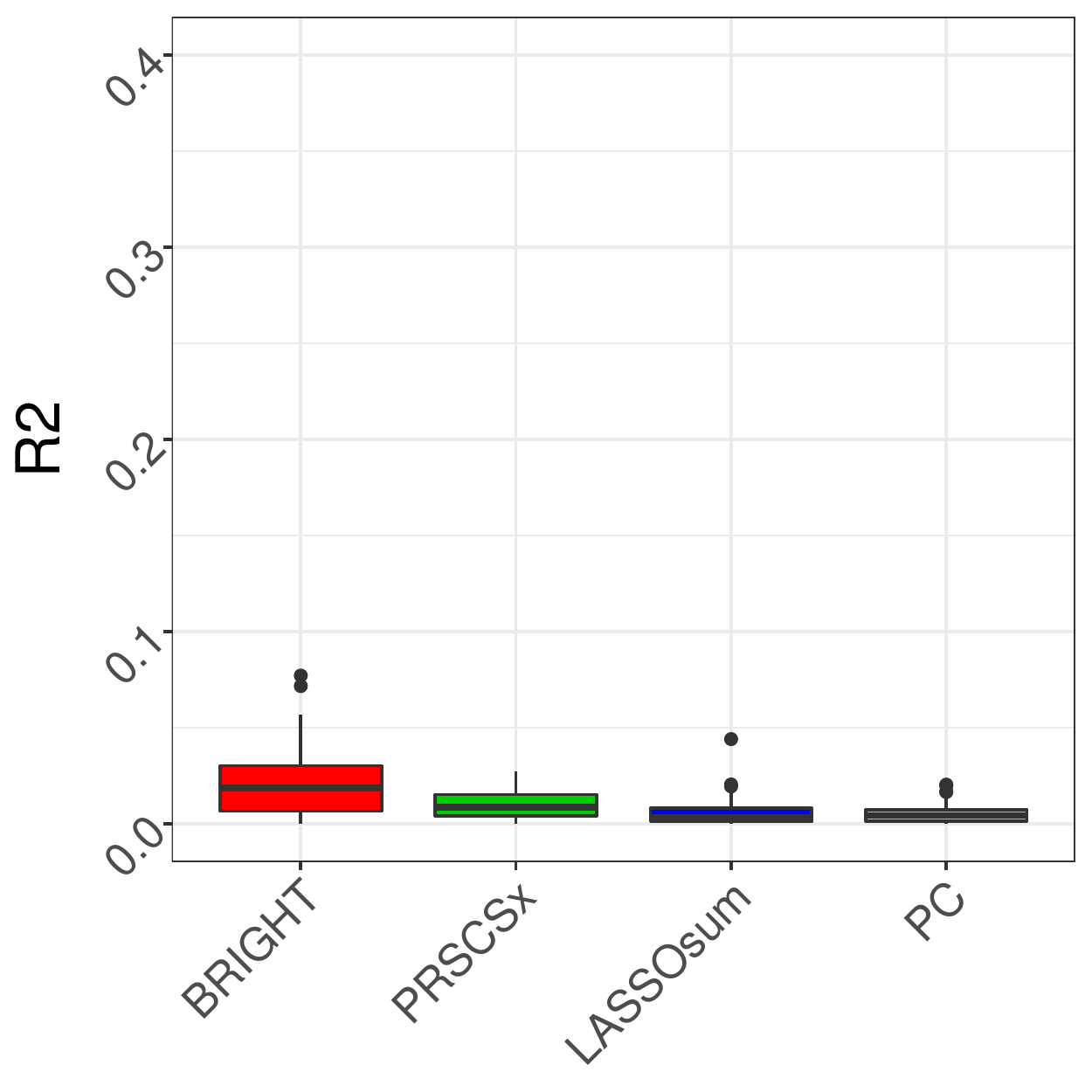}}
    \caption{Simulation results for high heterogeneous Caucasian and South Asian genetic effects, $\rho=0.7$. $h^2$ represents the heritability or signal to noise ratio and $|\mathcal{S}_0|$ represents the polygenicity level. Simulation figures from left to right are associated with a decreasing signal to noise ratio simulation setting and from up to down figures are having larger polygenicity levels.}
    \label{fig:rho=0.7}
\end{figure}

Figure \ref{fig:rho=1} shows the results  for homogeneous genetic effects  $\rho=1$. Compared with competing methods, the proposed BRIGHT estimation procedure consistently achieves the best prediction accuracy with regard to the $R^2$ criteria. Specifically, compared to PRSCSx, which does not assume a sparsity structure on the true effects, the BRIGHT method has significantly higher prediction performance. LASSOsum models are fitted on the Caucasian data only and then applied for the prediction of the South Asian population data; the results show that the naive trans-ethnic usage has a much worse prediction performance than the BRIGHT. Figures \ref{fig:rho=0.9} and \ref{fig:rho=0.7} show the results for medium ($\rho=0.9$) and high ($\rho=0.7$) heterogeneous genetic effects across populations. The overall pattern is similar to Figure \ref{fig:rho=1}. 

\subsection{Variable selection evaluation}

To evaluate the variable selection performance, we set the non-zero effect index set as $\mathcal{S}_0=\{1000,2000,3000,\dots,20000\}$ with  20 non-zero effects  $\beta_{0}^{\mathcal{S}_0}=\beta_{c}^{\mathcal{S}_0}=(1,-1,1,\dots,-1)^\top.$
The heritability and effect size heterogeneity are set to $h^2=0.1$ and $\rho=1$, respectively.

\begin{figure}[h]
    \centering
    \includegraphics[width=1\textwidth]{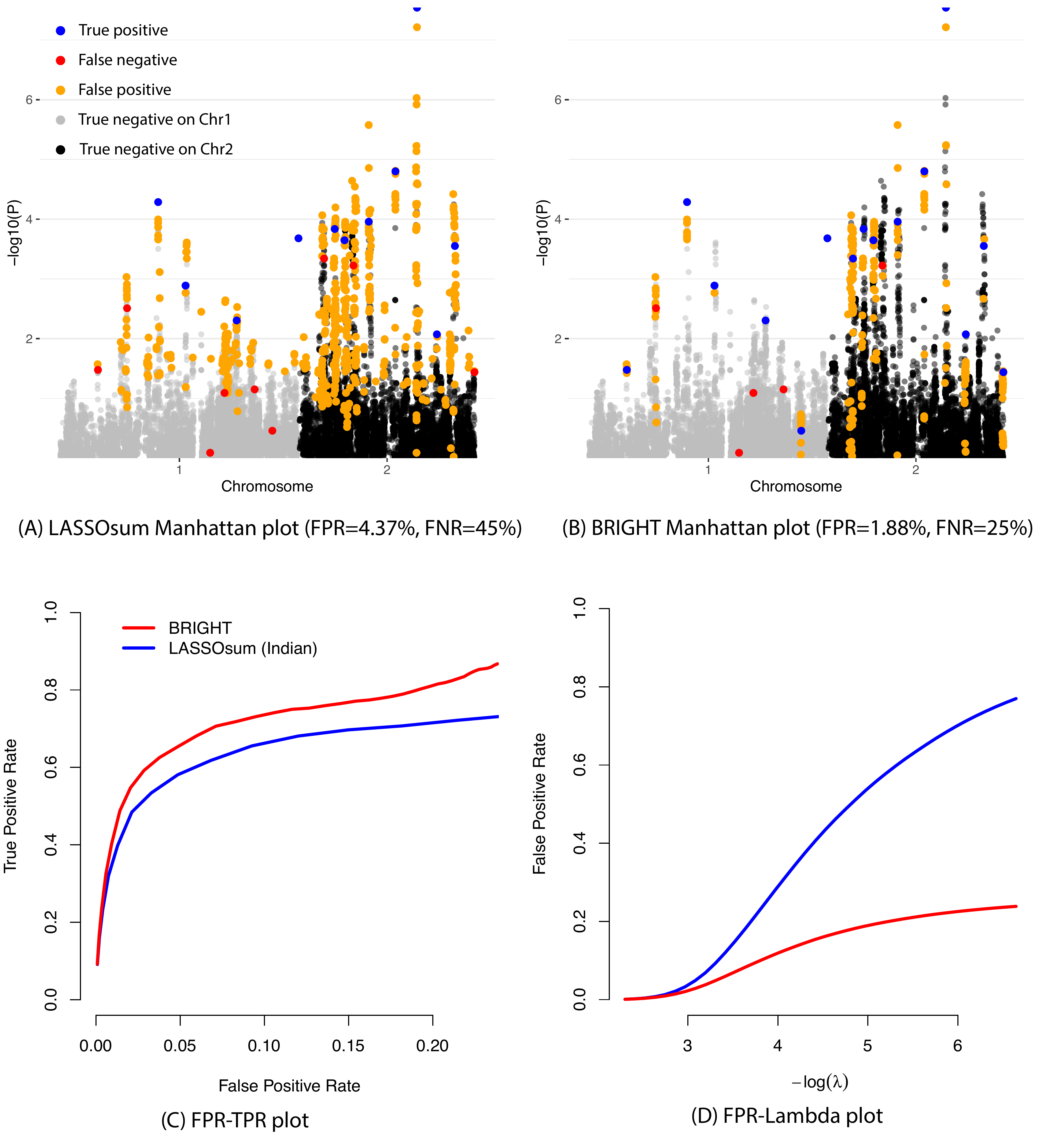}
    \caption{(A) and (B): Manhattan plots of the marginal correlation test p-value for each predictor-outcome pair based on South Asian GWAS data, showing variable selection results based on LASSOsum (A) and BRIGHT (B). Blue dots represent the true positive selections, red dots represent the false negative selections, yellow dots represent the false positive selections, and gray and black dots represent the true negative selections for predictors on chromosomes 1 and 2, respectively. (C) The variable selection true positive rate and false positive rate relationship for LASSOsum (blue) and BRIGHT (red) methods. (D) The variable selection false positive rate and $\lambda$ relationship for LASSOsum (blue) and BRIGHT (red).}
    \label{fig:Selection freq & FPR-TPR}
\end{figure}

Figure \ref{fig:Selection freq & FPR-TPR} (A) and (B) shows the Manhattan plots, where marginal correlation test p-values for
each predictor-outcome pair based on South Asian GWAS data are plotted. Comparing (A) and (B), we can see that the BRIGHT procedure has a larger power of identifying non-zero effects (blue dots) without misidentifying many zero effects (yellow dots). It achieves a False Positive Rate (FPR) of $1.88\%$ and False Negative Rate (FNR) of $25\%$, both lower than those in LASSOsum results ($4.37\%$ and $45\%$ respectively). (C) shows the True Positive Rate (TPR) and FPR for the BRIGHT (red) and LASSOsum (blue). The FPR-TPR curve of BRIGHT is always on top of the LASSOsum curve, indicating that BRIGHT has more accurate variable selection accuracy. (D) shows the FPR-$\lambda$ relationship for both methods. For a fixed $\lambda$, the BRIGHT FPR (red) is always smaller than the LASSOsum (blue), meaning that the former method is more robust in not including many false positive coefficients. 

The reason behind the above observations is that LASSOsum based only on South Asian data suffers from its limited sample size, so there is not enough power to identify true signals from noises. Therefore, the masked true signals from the many false positive selections in LASSOsum lead to impaired prediction performance. In contrast, the BRIGHT estimation procedure borrows information from the reference Caucasian population and the additional information helps BRIGHT to select more accurately the true signals.

\section{Real data applications}\label{Real data}

Psoriasis is a chronic, immune-mediated inflammatory skin disease with heritability ranging from 60\% to 90\%, according to a twin-study \citep{rahman2005genetic}. Although the high heritability illustrates a substantial genetic component in psoriasis, the genetic structures underlying the disease remain unclear. Most psoriasis GWASs are based on European origin and Chinese origin individuals, creating precision health inequities for other populations.

We apply the BRIGHT estimation procedure to reanalyze a South Asian psoriasis GWAS with the help of large sample-sized Caucasian psoriasis GWAS prior information. We reuse a 1,817 sample-sized South Asian psoriasis GWAS summary dataset from a previous study \citep{stuart2022transethnic} and treat it as the target dataset; we reuse the 11,675 sample-sized Caucasian GWAS from the same study as the prior information. Previous trans-ethnic meta-analysis research based on these datasets has found that effect sizes underlying psoriasis are concordant between Caucasian and South Asian populations (Spearman correlation 0.78) providing the potential opportunity of integrating information from the Caucasian GWAS to instruct and improve the model fitting of the South Asian population \citep{stuart2022transethnic}. Besides the above two training summary-level datasets, we also utilize an individual-level South Asian ethnic dataset consisting of 1644 psoriasis cases and 872 controls as independent testing data.

The two GWAS from South Asian and Caucasian populations consist of well-imputed markers ($R^2>0.7$); the common variants ($MAF\geq0.05$) with a call rate$<95\%$, the rare variants ($MAF<0.05$) with a call rate $<99\%$ and variants with a Hardy-Weinberg $p$ value smaller than $1\times10^{-6}$ were removed for quality control concerns \citep{stuart2022transethnic}. We first fit LASSOsum single-ethnic PRS on the Caucasian GWAS and create $\boldsymbol{\tilde\beta}$ as the prior information. Then, the BRIGHT estimation procedure is applied to integrate the South Asian GWAS and $\boldsymbol{\tilde\beta}$. In comparison, the PRSCSx is fitted on the South Asian and Caucasian GWAS, and the LASSOsum is also fitted solely on the South Asian GWAS.

\begin{figure}[h]
    \centering
    \includegraphics[width=1\textwidth]{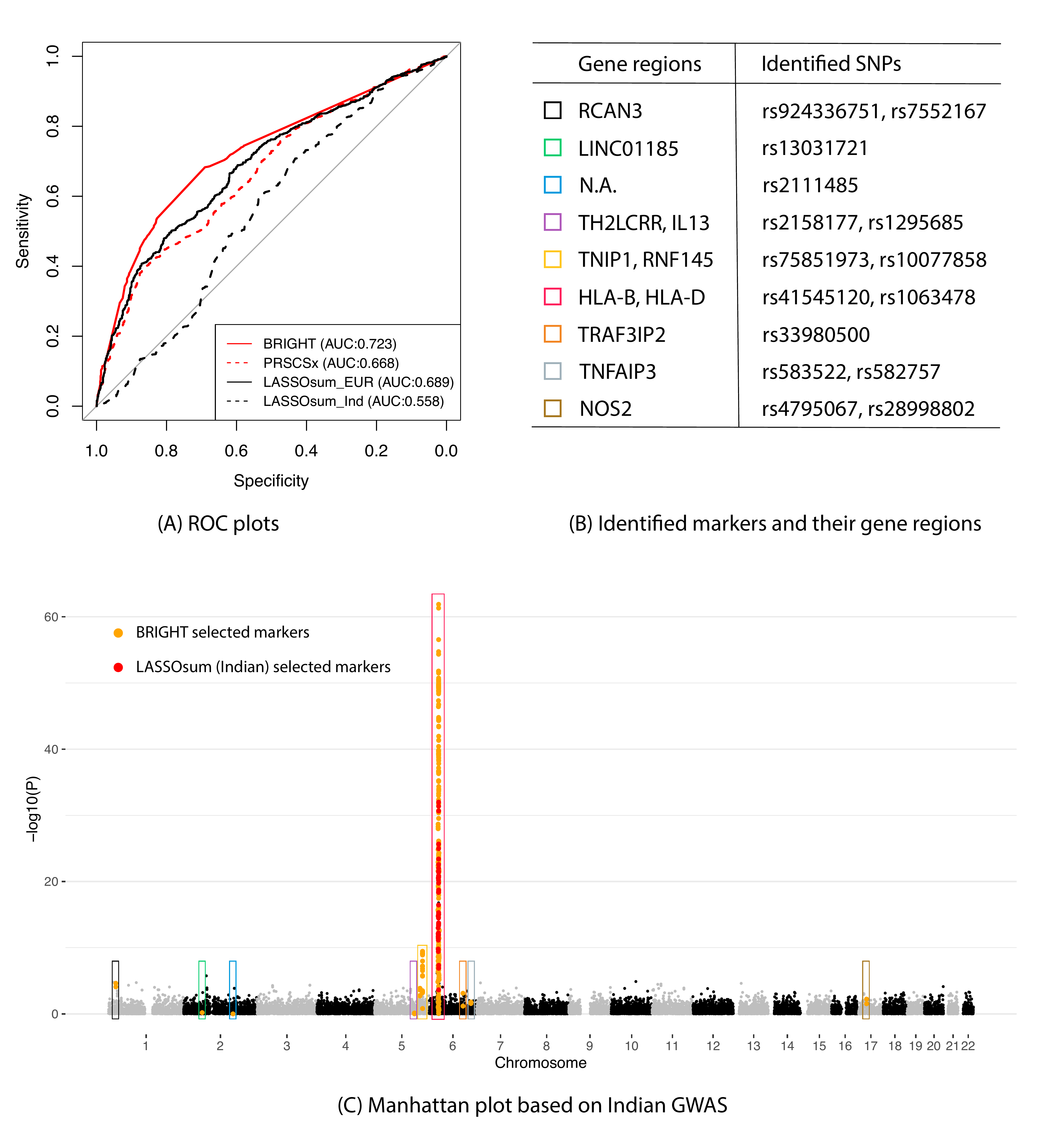}
    \caption{(A): The ROC curves and the AUC values evaluating the prediction performance of BRIGHT, PRSCSx, Caucasian-based LASSOsum and South Asian-based LASSOsum. (B): The table summarizing the identified SNPs and their corresponding gene regions, where only representative groups of SNPs are shown from each region. (C): The Manhattan plot of the marginal correlation test p-value for each marker-outcome pair based on South Asian GWAS data; the BRIGHT (orange) and South Asian-based LASSOsum (red) selected markers are highlighted among all markers. The different colored boxes correspond to the gene regions summarized in (B).}
    \label{fig:Real data}
\end{figure}

In the final results, the optimal $\eta$ is selected to be $\eta=0.12$; the relatively small $\eta$ indicates the potential genetic structure heterogeneity underlying psoriasis between South Asian and Caucasian populations. Nevertheless, the prediction performance comparison (see Figure \ref{fig:Real data} (A)) shows that a better South Asian model can be constructed by incorporating information from the Caucasian population; the proposed BRIGHT estimation procedure achieves the best prediction performance for the independent testing data (AUC=0.723). PRSCSx and Caucasian-based LASSOsum achieve close prediction performance (AUC=0.668 and 0.689), and the South Asian-based LASSOsum has the lowest performance (AUC=0.558).

Besides the prediction performance, variable selection results are also of great interest. In Figure \ref{fig:Real data} (B), we summarize the BRIGHT-identified gene regions, providing only representative groups of the identified SNPs in each region. We note that genes, LINC01185 \citep{villarreal2016candidate}, IL13 \citep{eder2011il13}, TNIP1 \citep{bowes2011confirmation}, RNF145 \citep{ming2021genetic}, genes in MHC region \citep{ozawa1988specific,tiilikainen1980psoriasis}, TRAF3IP2 \citep{huffmeier2010common}, TNFAIP3 \citep{tejasvi2012tnfaip3} and NOS2 \citep{garzorz2015nos2}, were previously reported with important contributions to the development of psoriasis. This concordance confirms the validity of our proposed BRIGHT method. In addition, the genes, RCAN3, TH2LCRR, and a SNP, rs2111485, not mapped to any gene region, are the newly identified markers, which deserve further biological investigation. Figure \ref{fig:Real data} (C) shows the Manhattan plot based on the South Asian GWAS, and the BRIGHT identified SNPs are highlighted in colored boxes corresponding to the gene region summary in Figure \ref{fig:Real data} (B). We also highlight the BRIGHT selected markers (orange) as well as the South Asian-based LASSOsum selected markers (red); we find that the South Asian-based LASSOsum not leveraging Caucasian information can only identify strong signals around HLA-B and HLA-D gene regions. In contrast, by borrowing information from the Caucasian population, the BRIGHT procedure identifies remarkably more signals, many of which have been confirmed in previous studies, as noted above.

\section{Discussion} \label{discussion}
The analysis of genetic data for minority racial groups has long suffered ``the curse of dimensionality''. To enhance the risk prediction of minority PRS with limited sample sizes, we propose a Bregman-divergence-based data integration estimation procedure borrowing information from well-studied majority populations. A unique challenge in our motivating setting is that due to data privacy, the individual data is not accessible to either the majority or the minority populations. To address these challenges, a Bregman-divergence is introduced to measure the information discrepancy and the resulting data integration procedure controls the relative weight of the external information,
identifying the most compatible ones and diminishing the weights of less relevant ones. The proposed method is computationally efficient and can be easily applied for ultrahigh-dimensional PRS construction and improve risk prediction for ethnic minority groups.

\newpage

\backmatter

\section*{Supplementary Materials}

\subsection*{Retrieve $r$ from GWAS summary statistics}\label{retrieve r}

GWAS aim to find marginally significant SNPs associated with the trait of interest. GWAS summary statistics usually include data sample size, coefficient estimates, t-statistics and p-values of marginal single SNP-trait association t-tests from Simple Linear Regressions (SLR),
$$\boldsymbol y=a_j+\boldsymbol{X}_{\cdot j} b_j+\epsilon.$$
Therefore, the standardized $\boldsymbol r_j$, $\boldsymbol r_j':=\frac{(\boldsymbol{X}_{\cdot j}-\boldsymbol{\overline X}_{\cdot j})^\top(\boldsymbol y-\boldsymbol{\overline{y}})}{\sqrt{(\boldsymbol{X}_{\cdot j}-\boldsymbol{\overline X}_{\cdot j})^\top(\boldsymbol{X}_{\cdot j}-\boldsymbol{\overline X}_{\cdot j})}\sqrt{(\boldsymbol y-\boldsymbol{\overline{y}})^\top(\boldsymbol y-\boldsymbol{\overline{y}})}}$ can be retrieved from the t-statistic and data sample size,
$$\boldsymbol r_j'=\frac{t}{\sqrt{t^2+n-2}}.$$
In the case, where only the GWAS summary statistic information is available, only the standardized $\boldsymbol r_j'$ can be retrieved for model fitting. As a result, the BRIGHT model can only be implemented on the standardized genotype matrix and standardized outcome,
which is sufficient for risk classification but cannot predict the outcome in the original scale. If researchers can also get access to the de-identified mean and variance estimate of each genotype, $\boldsymbol{\overline X}_{\cdot j}$ and $Var(\boldsymbol{X}_{\cdot j})$, and outcome, $\boldsymbol{\overline{y}}$ and $Var(\boldsymbol y)$, the original $\boldsymbol r_j:=\boldsymbol{X}_{\cdot j}^\top\boldsymbol y$ can be retrieved from the following equation:
$$\boldsymbol r_j=(n-1)\boldsymbol r_j'\sqrt{Var(\boldsymbol{X}_{\cdot j})Var(\boldsymbol y)}+n\boldsymbol{\overline X}_{\cdot j}\boldsymbol{\overline{y}}.$$
Then the original unstandardized BRIGHT model can be fitted.

\subsection*{Model implementation} \label{Model implementation}

The above BRIGHT objective function can be rewritten as below:
\begin{align}
    Q_{\eta,\lambda}(\boldsymbol\beta)&=\frac{1+\eta}{2}\boldsymbol{\beta}^\top \widetilde{\boldsymbol\Sigma}\boldsymbol{\beta}-\boldsymbol{\beta}^\top(\boldsymbol{r}+\eta  \widetilde{\boldsymbol\Sigma}\boldsymbol{\tilde\beta}) +\lambda||\boldsymbol{\beta}||_1, \label{eq: Final objective function}
\end{align}
which can be solved by extending the solution of the classic OLS LASSO problem. We follow the sub-differential approach to derive the solution of the BRIGHT estimation procedure. In the section below, we only provide essential intermediate steps. 

By taking sub-differential of $Q_{\eta,\lambda}(\boldsymbol\beta)$ we have,
\begin{equation}
  \frac{\partial{Q_{\eta,\lambda}(\bbeta)}}{\partial\bbeta_j}=\begin{cases}
    (1+\eta) \widetilde{\boldsymbol\Sigma}_{j}\boldsymbol{\beta}_{-j}+(1+\eta) \tilde\sigma_{jj} \beta_j-(\boldsymbol{r}_{j}+\eta  \widetilde{\boldsymbol\Sigma}_{j}\boldsymbol{\tilde \beta}) + \lambda sign(\bbeta_j),   & \text{if $\bbeta_j \neq 0$}\\
    (1+\eta) \widetilde{\boldsymbol\Sigma}_{j}\boldsymbol{\beta}_{-j}+(1+\eta) \tilde\sigma_{jj} \beta_j-(\boldsymbol{r}_{j}+\eta  \widetilde{\boldsymbol\Sigma}_{j}\boldsymbol{\tilde \beta}) + \lambda e,   & \text{if $\bbeta_j = 0$}
  \end{cases}
  \label{eq: derivative}
\end{equation}
where $e\in [-1,1]$, $\boldsymbol{\beta}_{-j}$ is the $\boldsymbol{\beta}$ coefficients with the $j^{th}$ element set to zero and $\widetilde{\boldsymbol\Sigma}_{j}$, $r_{j}$ are the $j^{th}$ row of the matrix $\widetilde{\boldsymbol\Sigma}$ and the $j^{th}$ element of vector $\boldsymbol r$, respectively.
Then the coordinate-like update for the $m^{th}$ iteration can be presented as
\begin{align*}
    \beta_j^{[m]}=ST(\frac{(\boldsymbol{r}_{j}+\eta  \widetilde{\boldsymbol\Sigma}_{j}\boldsymbol{\tilde \beta})}{(1+\eta)\tilde\sigma_{jj}}-\frac{ \widetilde{\boldsymbol\Sigma}_{j}}{\tilde\sigma_{jj}}(\beta_{1}^{[m]},\dots,\beta_{j-1}^{[m]},0,\beta_{j+1}^{[m-1]},\dots,\beta_{p}^{[m-1]})^\top,\frac{\lambda}{(1+\eta)\tilde\sigma_{jj}}),
\end{align*}
where $ST(\cdot,\cdot)$ is the soft-thresholding function. A good initialization of the above iterative algorithm can dramatically speed up the convergence time. For a single pair of $\eta$ and $\lambda$, due to no available information on the potential solution, we need to start from the null model, $\boldsymbol\beta^{[0]}=\boldsymbol 0$. However, in application, we usually need to evaluate a sequence of $\eta$ and $\lambda$ pairs for hyperparameter fine-tuning. In this case, when evaluating the current pair of hyperparameters, a warm start technique can be utilized to initialize the model on the solution of a pair of previously evaluated hyperparameters, which are close to the current ones. The algorithm will be stopped when the infinite norm of the difference between previous and current iterations estimates are smaller than the user specified criterion $\xi$.

For the hyperparameter fine-tuning we can either evaluate the model by utilizing the penalized BIC based on the training summary-level data or the correlation statistics between the predicted PRS score and the observed outcome based on an external test dataset. In either case it is important to discuss the plausible range of the two hyperparameters $\eta$ and $\lambda$. As discussed in Section \ref{BRIGHT estimation procedure}, $\eta$ is the weight adjusting the information to be incorporated from the Caucasian population; therefore, its range is from $0$ to $\infty$. $\lambda$ is the parameter adjusting the sparsity of coefficients; then, a maximum value corresponding to penalizing all coefficients to be 0 is available and can be presented as
\begin{align*}
    \lambda_{max}(\eta)=\max_{1\leq j\leq p}(\boldsymbol{r}_{j}+\eta  \widetilde{\boldsymbol\Sigma}_{j\cdot}\boldsymbol{\tilde \beta}).
\end{align*}
The plausible range for $\lambda$ is therefore $(0,\lambda_{max}(\eta)]$.

\subsection*{Proof of Theorem 1} \label{Proof}

We start from the basic inequality and rearrange its terms:
{
\begin{align}
    \frac{(1+\eta)}{2}\boldsymbol{\hat\beta}^\top \widetilde{\boldsymbol\Sigma}\boldsymbol{\hat\beta}-\boldsymbol{\hat\beta}^\top(\boldsymbol X^\top \boldsymbol{y}/n+\eta  \widetilde{\boldsymbol\Sigma}\boldsymbol{\tilde\beta}) +\lambda||\boldsymbol{\hat\beta}||_1&\leq\frac{(1+\eta)}{2}\boldsymbol{\beta}_0^\top \widetilde{\boldsymbol\Sigma}\boldsymbol{\beta}_0-\boldsymbol{\beta}_0^\top(\boldsymbol X^\top \boldsymbol{y}/n+\eta  \widetilde{\boldsymbol\Sigma}\boldsymbol{\tilde\beta}) +\lambda||\boldsymbol{\beta}_0||_1\nonumber\\
    \frac{1}{2}(\boldsymbol{\hat\beta}-\boldsymbol{\beta}_0)^\top\boldsymbol{\hat\Sigma}(\boldsymbol{\hat\beta}-\boldsymbol{\beta}_0)+\frac{\eta}{2}(\boldsymbol{\hat\beta}-\boldsymbol{\beta}_0)^\top\widetilde{\boldsymbol\Sigma}(\boldsymbol{\hat\beta}-\boldsymbol{\beta}_0)&\leq I_1+I_2+I_3+\lambda(||\boldsymbol\beta_0||_1-||\boldsymbol{\hat\beta}||_1), \label{Basic inequality}
\end{align}}
where
$$I_1=\left(\boldsymbol\epsilon^\top\boldsymbol X/n\right)(\boldsymbol{\hat\beta}-\boldsymbol{\beta}_0),\; I_2=\eta(\boldsymbol{\tilde\beta}-\boldsymbol{\beta}_0)^\top\widetilde{\boldsymbol\Sigma}(\boldsymbol{\hat\beta}-\boldsymbol{\beta}_0),\;I_3=\frac{1}{2}(\boldsymbol{\hat\beta}^\top\boldsymbol{\hat\Sigma}\boldsymbol{\hat\beta}-\boldsymbol{\beta}_0^\top\boldsymbol{\hat\Sigma}\boldsymbol{\beta}_0^+\boldsymbol{\beta}_0^\top\widetilde{\boldsymbol\Sigma}\boldsymbol{\beta}_0^-\boldsymbol{\hat\beta}^\top\widetilde{\boldsymbol\Sigma}\boldsymbol{\hat\beta}),$$ 
and $\boldsymbol{\hat\Sigma}=\frac{\boldsymbol X^\top\boldsymbol X}{n}$. In this proof, to simplify the notations we will use $\boldsymbol{\hat\beta}$ to replace $\boldsymbol{\hat\beta}_{\eta,\lambda}$.

In addition, we define the following event sets, 
$$\mathcal{F}_1:=\left\{\max_{1\leq j \leq p}|\boldsymbol\epsilon^\top\boldsymbol X_{j}|/n\leq\lambda/4\right\},$$
$$\mathcal{F}_2:=\left\{||\boldsymbol{\tilde\beta}-\boldsymbol\beta_0||_{\infty}\leq \sqrt{\frac{\log p}{n_c}}\right\},$$
$$\mathcal{F}_3:=\left\{||\widetilde{\boldsymbol\Sigma}-\boldsymbol\Sigma||_\infty \leq c_0(\frac{\log p}{\tilde n})^{(1-q)/2}\right\},$$
$$\mathcal{F}_4:=\left\{||\boldsymbol{\hat\Sigma}-\boldsymbol\Sigma||_\infty \leq (\frac{\log p}{n})^{1/2}\right\},$$
where $\lambda=A\left[\sqrt{\frac{logp}{n}}+\eta\sqrt{\frac{\log p}{n_c}}\left\{c_0(\frac{\log p}{\tilde n})^{(1-q)/2}+c_0\right\}\right]$. 
The following derivation will be restricted on $\mathcal{F}_1\bigcap\mathcal{F}_2\bigcap\mathcal{F}_3\bigcap\mathcal{F}_4$. We will show that the probability of $\mathcal{F}_1^c$ is tending to zero for large $n$ in the last part of this proof; probability of $\mathcal{F}_2^c$ tending to zero are assumed in Condition 7; probability of $\mathcal{F}_3^c$ tending to zero is given in Lemma 1; finally, the proof of probability of $\mathcal{F}_4^c$ tending to zero is given in Chapter 14 from \cite{buhlmann2011statistics}. Therefore, the probability of $\mathcal{F}_1\bigcap\mathcal{F}_2\bigcap\mathcal{F}_3\bigcap\mathcal{F}_4$ is tending to 1.
Then, we will investigate each and every $I$ defined above and bound them in probability.

Before that, we will review a lemma regarding random design in \cite{buhlmann2011statistics} and restate it as Lemma 2 below:

{\it \textbf{Lemma 2} Suppose the $\boldsymbol\Sigma$ compatibility condition holds and that $\max_{i,j}|\hat\Sigma_{ij}-\Sigma_{ij}|\leq\tilde\lambda$, then for all $\boldsymbol\delta$ satisfying $||\boldsymbol\delta^{\mathcal{S}_0^c}||_1\leq3||\boldsymbol\delta^{\mathcal{S}_0}||_1$ we have,
$$|\boldsymbol\delta^\top\boldsymbol{\hat\Sigma}\boldsymbol\delta-\boldsymbol\delta^\top\boldsymbol{\Sigma}\boldsymbol\delta|\leq16\tilde\lambda|\mathcal{S}_0|\boldsymbol\delta^\top\boldsymbol{\Sigma}\boldsymbol\delta/\phi_0^2$$

Therefore,
$$\boldsymbol\delta^\top\boldsymbol{\Sigma}\boldsymbol\delta-16\tilde\lambda|\mathcal{S}_0|\boldsymbol\delta^\top\boldsymbol{\Sigma}\boldsymbol\delta/\phi_0^2\leq\boldsymbol\delta^\top\boldsymbol{\hat\Sigma}\boldsymbol\delta\leq\boldsymbol\delta^\top\boldsymbol{\Sigma}\boldsymbol\delta+16\tilde\lambda|\mathcal{S}_0|\boldsymbol\delta^\top\boldsymbol{\Sigma}\boldsymbol\delta/\phi_0^2$$

Proof:
\begin{align*}
    |\boldsymbol\delta^\top\boldsymbol{\hat\Sigma}\boldsymbol\delta-\boldsymbol\delta^\top\boldsymbol{\Sigma}\boldsymbol\delta|&=|\boldsymbol\delta^\top(\boldsymbol{\hat\Sigma}-\boldsymbol{\Sigma})\boldsymbol\delta|\\
    &\leq \tilde\lambda||\boldsymbol\delta||_1^2\\
    &\leq \tilde\lambda(4||\boldsymbol\delta^{\mathcal{S}_0}||_1)^2\\
    &\leq 16\tilde\lambda|\mathcal{S}_0|\boldsymbol\delta^\top\boldsymbol{\Sigma}\boldsymbol\delta/\phi_0^2\square
\end{align*}
}

Similar property can be achieved for $\widetilde{\boldsymbol\Sigma}$, when the condition $\max_{i,j}|\widetilde\Sigma_{ij}-\Sigma_{ij}|\leq\tilde\lambda'$ is met. In the following proof $\tilde \lambda=(\frac{\log p}{n})^{1/2}$ and $\tilde \lambda'=c_0(\frac{\log p}{\tilde n})^{(1-q)/2}$ due to $\mathcal{F}_3$ and $\mathcal{F}_4$.

\begin{align}
    |I_1|&\leq||\boldsymbol\epsilon^\top\boldsymbol X/n||_\infty||\boldsymbol{\hat\beta}-\boldsymbol{\beta}_0||_1 \nonumber\\
    &=\left(\max_{1\leq j \leq p}|\boldsymbol\epsilon^\top\boldsymbol X_{j}|/n\right)||\boldsymbol{\hat\beta}-\boldsymbol{\beta}_0||_1 \nonumber\\
    &=\frac{\lambda}{4}||\boldsymbol{\hat\beta}-\boldsymbol{\beta}_0||_1,\label{I1}
\end{align}
where the first inequality is achieved from the definition of dual norm. 

\begin{align}
    |I_2|&\leq \eta||(\boldsymbol{\tilde\beta}-\boldsymbol{\beta}_0)^\top\widetilde{\boldsymbol\Sigma}||_\infty ||\boldsymbol{\hat\beta}-\boldsymbol{\beta}_0||_1 \nonumber\\
    &\leq \eta||(\boldsymbol{\tilde\beta}-\boldsymbol{\beta}_0)||_\infty
    ||\widetilde{\boldsymbol\Sigma}||_\infty ||\boldsymbol{\hat\beta}-\boldsymbol{\beta}_0||_1 \nonumber\\
    &\leq \eta||(\boldsymbol{\tilde\beta}-\boldsymbol{\beta}_0)||_\infty
    (||\widetilde{\boldsymbol\Sigma}-\boldsymbol\Sigma||_\infty+||\boldsymbol\Sigma||_\infty) ||\boldsymbol{\hat\beta}-\boldsymbol{\beta}_0||_1 \nonumber\\
    &\leq \eta\sqrt{\frac{\log p}{n_c}}
    \left\{c_0(\frac{\log p}{\tilde n})^{(1-q)/2}+c_0\right\} ||\boldsymbol{\hat\beta}-\boldsymbol{\beta}_0||_1 \nonumber\\
    &\leq\frac{\lambda}{4}||\boldsymbol{\hat\beta}-\boldsymbol{\beta}_0||_1, \label{I2}
\end{align}
where the first two inequalities are achieved from the definition of dual norm and matrix induced norm respectively; the third inequality is due to matrix induced infinite norm triangle inequality; the fourth inequality is achieved based on $\mathcal{F}_2$ and $\mathcal{F}_3$ and {\it \textbf{Condition 2}}.

\begin{align}
    |I_3|&= \frac{1}{2}\left\{\boldsymbol{\hat\beta}^\top(\boldsymbol{\hat\Sigma}-\boldsymbol\Sigma)\boldsymbol{\hat\beta}-\boldsymbol{\beta}_0^\top(\boldsymbol{\hat\Sigma}-\boldsymbol\Sigma)\boldsymbol{\beta}_0+\boldsymbol{\beta}_0^\top(\widetilde{\boldsymbol\Sigma}-\boldsymbol\Sigma)\boldsymbol{\beta}_0-\boldsymbol{\hat\beta}^\top(\widetilde{\boldsymbol\Sigma}-\boldsymbol\Sigma)\boldsymbol{\hat\beta}\right\} \nonumber\\
    &\leq \frac{\tilde\lambda+\tilde\lambda'}{2}(||\boldsymbol{\hat\beta}||_1^2-||\boldsymbol{\beta}_0||_1^2)+(\tilde\lambda+\tilde\lambda')||\boldsymbol{\beta}_0||_1^2 \nonumber\\
    &\leq \frac{8(\tilde\lambda+\tilde\lambda')|\mathcal{S}_0|}{\phi_0^2}(\boldsymbol{\hat\beta}-\boldsymbol{\beta}_0)^\top\boldsymbol{\Sigma}(\boldsymbol{\hat\beta}-\boldsymbol{\beta}_0)+(\tilde\lambda+\tilde\lambda')||\boldsymbol{\beta}_0||_1^2, \label{I3}
\end{align}
where the second inequality is due to {\it \textbf{Lemma 2}}; the third inequality is based on the triangle inequality of the quadratic norm with $\boldsymbol\Sigma$ being positive definite.

In addition, we also give the lower bound of the two terms on the left-hand side of (\ref{Basic inequality}) based on {\it \textbf{Lemma 2}}.
\begin{align}
    \frac{1}{2}(\boldsymbol{\hat\beta}-\boldsymbol{\beta}_0)^\top\boldsymbol{\hat\Sigma}(\boldsymbol{\hat\beta}-\boldsymbol{\beta}_0)\geq\frac{\phi_0^2-16\tilde\lambda|\mathcal{S}_0|}{2\phi_0^2}(\boldsymbol{\hat\beta}-\boldsymbol{\beta}_0)^\top\boldsymbol{\Sigma}(\boldsymbol{\hat\beta}-\boldsymbol{\beta}_0)\label{BI1}
\end{align}
\begin{align}
    \frac{\eta}{2}(\boldsymbol{\hat\beta}-\boldsymbol{\beta}_0)^\top\widetilde{\boldsymbol\Sigma}(\boldsymbol{\hat\beta}-\boldsymbol{\beta}_0)\geq\frac{\eta(\phi_0^2-16\tilde\lambda'|\mathcal{S}_0|)}{2\phi_0^2}(\boldsymbol{\hat\beta}-\boldsymbol{\beta}_0)^\top\boldsymbol{\Sigma}(\boldsymbol{\hat\beta}-\boldsymbol{\beta}_0)\label{BI2}
\end{align}

Then replacing each term in (\ref{Basic inequality}) to their bound in (\ref{I1}), (\ref{I2}), (\ref{I3}), (\ref{BI1}) and (\ref{BI2}) we have:
{\small
\begin{align}
    \left\{\frac{1+\eta}{2}-\frac{32\tilde\lambda|\mathcal{S}_0|+16(1+\eta)\tilde\lambda'|\mathcal{S}_0|}{2\phi_0^2}\right\}||\boldsymbol{\hat\beta}-\boldsymbol{\beta}_0||_{2,\boldsymbol\Sigma}^2&\leq\frac{\lambda}{2}||\boldsymbol{\hat\beta}-\boldsymbol{\beta}_0||_1+\lambda(||\boldsymbol\beta_0||_1-||\boldsymbol{\hat\beta}||_1)+(\tilde\lambda+\tilde\lambda')||\boldsymbol{\beta}_0||_1^2\nonumber\\
    \left\{\frac{1+\eta}{2}-\frac{32\tilde\lambda|\mathcal{S}_0|+16(1+\eta)\tilde\lambda'|\mathcal{S}_0|}{2\phi_0^2}\right\}||\boldsymbol{\hat\beta}-\boldsymbol{\beta}_0||_{2,\boldsymbol\Sigma}^2&\leq2\lambda||\boldsymbol{\hat\beta}^{\mathcal{S}_0}-\boldsymbol{\beta}_0^{\mathcal{S}_0}||_1+(\tilde\lambda+\tilde\lambda')||\boldsymbol{\beta}_0||_1^2\nonumber\\
    \left\{\frac{1+\eta}{2}-\frac{32\tilde\lambda|\mathcal{S}_0|+16(1+\eta)\tilde\lambda'|\mathcal{S}_0|}{2\phi_0^2}\right\}||\boldsymbol{\hat\beta}-\boldsymbol{\beta}_0||_{2,\boldsymbol\Sigma}^2&\leq2\lambda\frac{\sqrt{|\mathcal{S}_0|}||\boldsymbol{\hat\beta}-\boldsymbol{\beta}_0||_{2,\boldsymbol\Sigma}}{\phi_0}+(\tilde\lambda+\tilde\lambda')||\boldsymbol{\beta}_0||_1^2\nonumber\\
    \left\{\frac{1+\eta}{2}-\frac{32\tilde\lambda|\mathcal{S}_0|+16(1+\eta)\tilde\lambda'|\mathcal{S}_0|}{2\phi_0^2}\right\}||\boldsymbol{\hat\beta}-\boldsymbol{\beta}_0||_{2,\boldsymbol\Sigma}^2&\leq\frac{b\lambda^2{|\mathcal{S}_0|}}{\phi_0^2}+\frac{1}{b}||\boldsymbol{\hat\beta}-\boldsymbol{\beta}_0||_{2,\boldsymbol\Sigma}^2+(\tilde\lambda+\tilde\lambda')||\boldsymbol{\beta}_0||_1^2\nonumber\\
    \left\{\frac{1+\eta}{2}-\frac{32\tilde\lambda|\mathcal{S}_0|+16(1+\eta)\tilde\lambda'|\mathcal{S}_0|}{2\phi_0^2}-\frac{1}{b}\right\}||\boldsymbol{\hat\beta}-\boldsymbol{\beta}_0||_{2,\boldsymbol\Sigma}^2&\leq\frac{b\lambda^2{|\mathcal{S}_0|}}{\phi_0^2}+(\tilde\lambda+\tilde\lambda')||\boldsymbol{\beta}_0||_1^2\nonumber\\
    \left\{\frac{(b+\eta b-2)\phi_0^2-\{32\tilde\lambda|\mathcal{S}_0|+16(1+\eta)\tilde\lambda'|\mathcal{S}_0|\}b}{2\phi_0^2b}\right\}||\boldsymbol{\hat\beta}-\boldsymbol{\beta}_0||_{2,\boldsymbol\Sigma}^2&\leq\frac{b\lambda^2{|\mathcal{S}_0|}}{\phi_0^2}+(\tilde\lambda+\tilde\lambda')||\boldsymbol{\beta}_0||_1^2\nonumber\\
    ||\boldsymbol{\hat\beta}-\boldsymbol{\beta}_0||_{2,\boldsymbol\Sigma}^2&\leq\frac{2b^2\lambda^2{|\mathcal{S}_0|}+2\phi_0^2b(\tilde\lambda+\tilde\lambda')||\boldsymbol{\beta}_0||_1^2}{(b+\eta b-2)\phi_0^2-\{32\tilde\lambda|\mathcal{S}_0|+16(1+\eta)\tilde\lambda'|\mathcal{S}_0|\}b}\label{final1}
\end{align}}

Let $b=\frac{3}{1+\eta}$ and divide $\phi_0^2$ for both the denominator and numerator on the right-hand side, we further write (\ref{final1}) as:
\begin{align}
    ||\boldsymbol{\hat\beta}-\boldsymbol{\beta}_0||_{2,\boldsymbol\Sigma}^2&\leq\frac{18\lambda^2{|\mathcal{S}_0|}/\{(1+\eta)^2\phi_0^2\}+6(\tilde\lambda+\tilde\lambda')||\boldsymbol{\beta}_0||_1^2/(1+\eta)}{1-\{96\tilde\lambda|\mathcal{S}_0|/\phi_0^2+48(1+\eta)\tilde\lambda'|\mathcal{S}_0|/\phi_0^2\}/(1+\eta)}\nonumber\\
    &\leq\frac{18|\mathcal{S}_0|}{\phi_0^2 B}\frac{\lambda^2}{(1+\eta)^2}+\frac{6(\tilde\lambda+\tilde\lambda')||\boldsymbol{\beta}_0||_1^2}{B(1+\eta)},\label{final2}
\end{align}
where $B=1-\{96\tilde\lambda|\mathcal{S}_0|/\phi_0^2+48(1+\eta)\tilde\lambda'|\mathcal{S}_0|/\phi_0^2\}/(1+\eta)$.

As a result, plug in $\lambda=A\left[\frac{\sqrt{logp}}{\sqrt{n}}+\eta\frac{\sqrt{\log p}}{\sqrt{n_c}}
    \left\{c_0(\frac{\log p}{\tilde n})^{(1-q)/2}+c_0\right\}\right]$ (\ref{final2}) is reduced to

\begin{align}
        ||\boldsymbol{\hat\beta}-\boldsymbol{\beta}_0||_{2,\boldsymbol\Sigma}^2&\leq\frac{18|\mathcal{S}_0|}{\phi_0^2B}\frac{\lambda^2}{(1+\eta)^2}+\frac{6(\tilde\lambda+\tilde\lambda')||\boldsymbol{\beta}_0||_1^2}{B(1+\eta)}\nonumber\\
        &\leq\frac{36A^2|\mathcal{S}_0|}{\phi_0^2B}\left[\frac{1}{(1+\eta)^2}\frac{{logp}}{{n}}+\frac{\eta^2}{(1+\eta)^2}\frac{{\log p}}{{n_c}}\left\{c_0(\frac{\log p}{\tilde n})^{(1-q)/2}+c_0\right\}^2\right]+\frac{6(\tilde\lambda+\tilde\lambda')||\boldsymbol{\beta}_0||_1^2}{B(1+\eta)}.\label{final3}
\end{align}

{\it Proof of probability of $\mathcal{F}_1^c$ tends to zero}
\begin{align*}
    Pr(\mathcal{F}_1^c)&=Pr\left (\max_{1\leq j \leq p}|2\boldsymbol\epsilon^\top\boldsymbol X_{j}|/n>\frac{\lambda}{4}\right )\\
    &\leq Pr\left (\max_{1\leq j \leq p}|2\boldsymbol\epsilon^\top\boldsymbol X_{j}|>A\frac{\sqrt{nlogp}}{4}\right )\\
    &\leq \sum_{j=1}^p Pr\left (|\boldsymbol\epsilon^\top\boldsymbol X_{j}|>A\frac{\sqrt{nlogp}}{8}\right )\\
\end{align*}
Bernstein's inequality condition
\begin{align*}
    E(|\epsilon_{i}\boldsymbol X_{ij}|^m)&\leq E(|\epsilon_{i}|^m)|K|^m\\
    &\leq m!(MK)^m\\
    &=m!(MK)^{m-2}2(MK)^{2}/2
\end{align*}
By applying Bernstein inequality, we have:
\begin{align*}
    Pr(\mathcal{F}_1^c)&\leq 2pexp\left\{-\frac{1}{2}\frac{(A^2/64)nlogp}{n2(MK)^{2}+(MK)(A\sqrt{nlogp}/8)}\right\}\\
    &\leq 2pexp\left\{-\frac{1}{2}\frac{(A^2/64)nlogp}{M' (n+A\sqrt{nlogp}/8)}\right\}\\
    &\leq 2exp\left\{-\frac{1}{2}\frac{(A^2/64)nlogp}{M' (n+A\sqrt{nlogp}/8)}+logp\right\}\\
    &\leq 2exp\left\{-\frac{1}{2}\frac{(A^2/64-2M')nlogp-M'(A/4)\sqrt{n}(logp)^{3/2}}{M' (n+A\sqrt{nlogp}/8)}\right\}\\
    &\leq 2exp\left\{-\frac{1}{2}\frac{(A^2/64-2M')logp-M'(A/4)(logp)^{3/2}/\sqrt{n}}{M' (1+A\sqrt{logp}/(8\sqrt{n}))}\right\}\\
\end{align*}
where, $M'=max\{MK,2(MK)^{2}\}$. It is noticed that when $A$ is large enough and $logp=o(n)$ then $Pr(\mathcal{F}_1^c)\rightarrow0$. $\square$

\bibliographystyle{biom}
\bibliography{ref}

\label{lastpage}

\end{document}